\PassOptionsToPackage{table,xcdraw}{xcolor}
\documentclass[11pt,a4paper]{article}
\pdfoutput=1

\makeatletter
\def\@fpheader{}
\makeatother

\usepackage{jheppub}
\usepackage{gensymb}
\DeclareSymbolFont{matha}{OML}{txmi}{m}{it}% txfonts
\DeclareMathSymbol{\varv}{\mathord}{matha}{118}
\usepackage{amsmath}
\usepackage{amssymb}
\usepackage{mathrsfs}
\usepackage{graphicx}
\usepackage{subfigure}
\usepackage{pstricks}
\usepackage{bm}
\usepackage{pbox}
\usepackage{placeins}
\usepackage{booktabs}
\usepackage[T1]{fontenc}
\usepackage{footnote}
\usepackage{pdfpages}
\usepackage{hhline}
\usepackage{multirow}
\usepackage{multicol}
\usepackage{enumitem}
\usepackage[toc,page]{appendix}
\allowdisplaybreaks
\usepackage{comment}
\usepackage{float}
\usepackage{slashed}
\usepackage{xcolor}
\usepackage{textcomp}
\usepackage{multirow}
\usepackage[numbers]{natbib}
\usepackage{notoccite}
\usepackage{indentfirst}
\usepackage{jheppub}
\usepackage{gensymb}
\DeclareSymbolFont{matha}{OML}{txmi}{m}{it}% txfonts
\DeclareMathSymbol{\varv}{\mathord}{matha}{118}
\usepackage{amsmath}
\usepackage{array}
\usepackage{amssymb}
\usepackage{graphicx}
\usepackage{subfigure}
\usepackage{pstricks}
\usepackage{bm}
\usepackage{pbox}
\usepackage{placeins}
\usepackage{booktabs}
\usepackage[T1]{fontenc}
\usepackage{footnote}
\usepackage{pdfpages}
\usepackage{hhline}
\usepackage{multirow}
\usepackage{multicol}
\usepackage{enumitem}
\usepackage[toc,page]{appendix}
\allowdisplaybreaks
\usepackage{comment}
\usepackage{float}
\usepackage{slashed}
\usepackage{xcolor}
\usepackage{textcomp}
\usepackage{multirow}
\usepackage[numbers]{natbib}
\usepackage{notoccite}
\usepackage{soul}
\usepackage{extarrows}
\usepackage{makecell}
\usepackage{longtable}
\usepackage{physics}
\usepackage{comment}
\usepackage{blindtext}
\usepackage{csquotes}
\usepackage{soul}
\usepackage{indentfirst}
\DeclareUnicodeCharacter{0304}{ }
\usepackage{rotating}
\usepackage{tabularx}
\usepackage{extarrows}
\usepackage{physics}
\usepackage{comment}
\usepackage{cancel}
\usepackage{blindtext}
\usepackage{csquotes}
\usepackage[none]{hyphenat}
\DeclareUnicodeCharacter{0304}{ }
\usepackage{rotating}
\usepackage{tabularx}
\usepackage[many]{tcolorbox}
\definecolor{fg}{RGB}{34,139,34}

\renewcommand{\thetable}{\Roman{table}}

\def\figureautorefname~#1\null{Fig.\,#1\null}
\def\equationautorefname~#1\null{Eq.\,(#1)\null}
\def\tableautorefname~#1\null{Tab.\,#1\null}
\definecolor{MyDarkBlue}{rgb}{0.1, 0.1, 0.8} 
\definecolor{MyLightBlue}{rgb}{0.22,0.51,0.9}
\definecolor{MyGreen}{rgb}{0.0, 0.5, 0.0}
\definecolor{BrickRed}{rgb}{0.8, 0.25, 0.33}

\RequirePackage{hyperref}
\hypersetup{colorlinks=true, citecolor=MyGreen,linkcolor=BrickRed, urlcolor=MyLightBlue}
\usepackage{cleveref}
\crefname{equation}{Eq.}{Eqs.} % capitalize "E", no period
\crefrangelabelformat{equation}{(#3#1#4--#5#2#6)}
\title{\bf Quantum Tomography and Entanglement in Semi-Leptonic $h\to VV^*$ Decays at Higher Orders} 
\author[a]{Dorival Gon\c{c}alves,}
\author[a,b]{Ajay Kaladharan,}
\author[c]{Alberto Navarro}
\affiliation[a]{Department of Physics, Oklahoma State University, Stillwater, OK 74078, USA}
\affiliation[b]{International Centre for Theoretical Physics Asia-Pacific, University of Chinese Academy of Sciences, 100190 Beĳing, China}
\affiliation[c]{Institute for Convergence of Basic Studies, Seoultech, Seoul 01811, Republic of Korea}
\emailAdd{dorival@okstate.edu}\emailAdd{kaladharan.ajay@ucas.ac.cn}\emailAdd{alberto.navarro1001@seoultech.ac.kr}

\abstract{Angular correlations in Higgs decays to electroweak gauge bosons, $h \to ZZ^*, WW^*$, provide a powerful probe of both new physics effects and quantum information observables. We present a systematic study of semi-leptonic Higgs decays $h \to V V^* \to \ell^+\ell^- q\bar{q}$ and $\ell^\pm \nu_\ell q\bar{q}'$, including finite final state fermion masses, NLO QCD, and NLO electroweak corrections. We show that finite final state quark masses can induce effects that go beyond the two-qutrit description in more inclusive regimes, while remaining controllable with suitable kinematic selections. QCD corrections lead to modest percent-level shifts, whereas electroweak corrections can significantly modify the angular structure, particularly in the $h\to ZZ^*$ channels. We assess the impact of these effects on the reconstructed density matrix and entanglement measures, finding that, although they modify the angular observables, the semi-leptonic channels are better approximated by a two-qutrit description than the fully leptonic decay $h\to e^+e^-\mu^+\mu^-$.}

\begin{document}

\maketitle

\begin{sloppypar}

%%%%%%%%%%%%%%%%%%%%%%%%%%%%%%%%%%%%%%%%%%%%%%%%%%%%
\section{Introduction}
\label{sec:intro}

The decays of the Higgs boson into electroweak gauge bosons, $h\to ZZ^*, WW^*$, are a cornerstone of the physics program at the Large Hadron Collider (LHC). These channels provide some of the most sensitive probes of the Higgs sector, enabling detailed tests of the Standard Model (SM) and offering direct access to possible new physics effects~\cite{ATLAS:2012yve,CMS:2012qbp,Buszello:2002uu,Godbole:2007cn,Bolognesi:2012mm,Englert:2012xt,Artoisenet:2013puc,Caola:2013yja,Campbell:2013una,Englert:2014aca,Azatov:2014jga,Buschmann:2014sia,Corbett:2015ksa,Goncalves:2017gzy,Brehmer:2017lrt,Goncalves:2017iub,Goncalves:2018pkt,CMS:2019ekd,Gritsan:2020pib,Goncalves:2020vyn,Hall:2022bme,CMS:2022ley,ATLAS:2023dnm,ATLAS:2023mqy,Bhardwaj:2024lyr,Haisch:2025jqr}. Their multi-body kinematics give rise to a rich set of angular observables that encode detailed information about the tensor structure of the Higgs couplings and offer a powerful probe of deviations from the SM, particularly in searches for CP-violating interactions and within effective field theory interpretations. A precise understanding of their theoretical description, including the impact of radiative corrections, is therefore crucial for both precision measurements and new physics searches.

More recently, these observables have also been reinterpreted within the framework of quantum information~\cite{Barr:2021zcp,Aguilar-Saavedra:2022wam,Ashby-Pickering:2022umy,Aguilar-Saavedra:2022mpg,Aoude:2023hxv,Fabbrichesi:2023cev,Fabbrichesi:2023jep,Fabbri:2023ncz,Bernal:2023ruk,Morales:2023gow,Bi:2023uop,Barr:2024djo,Aguilar-Saavedra:2024whi,Subba:2024mnl,Bernal:2024xhm,Sullivan:2024wzl,Aguilar-Saavedra:2024jkj,Grossi:2024jae,Ruzi:2024cbt,Wu:2024ovc,Low:2025aqq,Ding:2025mzj,DelGratta:2025qyp,Goncalves:2025mvl,Goncalves:2025xer,De:2025dpo,Ruzi:2025jql,Gu:2025ijz,CMS:2025anw,Aguilar-Saavedra:2025njw,Pelliccioli:2026ltl,Gabrielli:2026tnl}. In this perspective, the angular distributions enable the reconstruction of the spin density matrix of the diboson system via quantum tomography, providing a complete characterization of its polarization and spin correlations, granting access to intrinsically quantum features, most notably entanglement, in high-energy collider processes. The feasibility of such measurements has been demonstrated experimentally. In particular, the $h\to ZZ^*\to 4\ell$ channel has been used by ATLAS to establish entangled diboson configurations at the $4.7\sigma$ level~\cite{ATLAS:2026hye}. This study builds on earlier observations of entanglement in $t\bar t$ production at TeV energies~\cite{Afik:2020onf,Fabbrichesi:2021npl,Severi:2021cnj,Aguilar-Saavedra:2022uye,Afik:2022kwm,Afik:2022dgh,Severi:2022qjy,Aoude:2022imd,Dong:2023xiw,Aguilar-Saavedra:2023hss,Han:2023fci,Sakurai:2023nsc,Cheng:2023qmz,Maltoni:2024tul,Maltoni:2024csn,White:2024nuc,Dong:2024xsg,Dong:2024xsb,Cheng:2024btk,Altomonte:2024upf,Han:2024ugl,Cheng:2025cuv,Afik:2025ejh,Nason:2025hix,ATLAS:2023fsd,CMS:2024pts,CMS:2024zkc,Afik:2025grr,Guo:2026yhz, Fang:2026ddi,Lin:2025eci, Durupt:2025wuk,Choi:2026omc,Altakach:2026fpl}. Taken together, these measurements demonstrate that collider data can probe uniquely quantum features of fundamental interactions at the TeV scale.\footnote{Quantum tomography provides a systematic framework to reconstruct the full quantum state of particle systems, making it possible to identify structures that are not directly visible in conventional kinematic observables. For instance, recent experimental studies have reported features consistent with a short-lived quasi-bound state of $t\bar t$ near threshold~\cite{CMS:2025kzt}.}

A central ingredient in this program is the mapping between the measured angular distributions and the underlying density matrix, which relies on the assumption that the intermediate gauge bosons can be treated as effective spin-1 degrees of freedom. Under this assumption, the $VV^*$ system can be described as a bipartite spin-1 system, corresponding to a two-qutrit state. However, this picture is not exact in realistic Higgs decays. For $m_h\simeq 125$~GeV, at least one of the vector bosons is off-shell, and additional effects, such as finite final state fermion masses and higher-order QCD and electroweak corrections, can introduce contributions beyond the pure spin-1 description. These effects can modify the angular structure and, in certain regimes, challenge the consistency of the two-qutrit interpretation.

This issue has been explored in fully leptonic final states. In the $h\to e^+e^-\mu^+\mu^-$ channel, next-to-leading order (NLO) electroweak corrections induce sizable effects on the angular distributions, signaling a breakdown of the effective two-qutrit description~\cite{Grossi:2024jae,DelGratta:2025qyp,Goncalves:2025mvl,DelGratta:2025xjp}. In contrast, the $h \to \ell^+ \nu_\ell \ell'^- \bar{\nu}_{\ell'}$ channel exhibits a more stable behavior under higher-order effects~\cite{Goncalves:2025xer}, largely due to the chiral structure of the $W$ couplings. These differences highlight that higher-order corrections go beyond a quantitative refinement of the leading-order picture, as they can directly impact the interpretation of the underlying quantum state. Accordingly, these effects should be understood and consistently incorporated, either within the signal modeling or in the definition of the background~\cite{Aguilar-Saavedra:2025byk,Aguilar-Saavedra:2026wuq}.

Semi-leptonic Higgs decays, $h\to \ell^+\ell^- q\bar q$ and $h\to \ell^\pm\nu_\ell q\bar q'$, provide a natural next step in this program. They combine the clean experimental signature of leptonic final states with the larger branching fractions of hadronic modes. Recent phenomenological studies at leading order indicate that the $h\to WW^*\to \ell^\pm\nu_\ell q\bar q'$ mode offers promising sensitivity to quantum correlations, with evidence for Bell inequality violation achievable with integrated luminosities of order $300~\text{fb}^{-1}$ and observation at the high-luminosity LHC~\cite{Fabbri:2023ncz}. In combination with the recent ATLAS measurement in the $h\to 4\ell$ channel~\cite{ATLAS:2026hye}, these results illustrate both the feasibility and the strong motivation to extend the analysis to semi-leptonic final states.

In this work, we present a systematic study of finite final state mass effects and higher-order QCD and electroweak corrections to angular observables in semi-leptonic $h\to ZZ^*, WW^*$ decays. We derive the angular coefficients at leading order and quantify the impact of finite fermion masses. We then include NLO QCD corrections and assess their effect on the angular structure and the reconstruction of the spin density matrix. Finally, we incorporate NLO electroweak corrections and evaluate their impact on both the angular observables and the associated quantum information observables. Throughout, we test the stability of the reconstructed density matrix and the validity of the two-qutrit description. The resulting density matrix reconstructions should be interpreted as parton level benchmarks for future experimental analyses.

Beyond their role in entanglement studies, the angular observables considered here provide a model-independent interface between theory and experiment, where multipole expansions offer a systematic basis for new physics analyses and a direct probe of the underlying dynamics~\cite{ATLAS:2016rnf,ATLAS:2018gqq,Goncalves:2018fvn,Goncalves:2018ptp,ATLAS:2019zrq,CMS:2019nrx,MammenAbraham:2022yxp,Bhardwaj:2023ufl,Denner:2023ehn,Denner:2024tlu,CMS:2024ony,Carrivale:2025mjy,Grossi:2024jae,DelGratta:2025qyp,Goncalves:2025mvl,Goncalves:2025xer}. As the LHC advances toward its high-luminosity phase, with substantially larger data sets, the precise characterization of these angular correlations will become increasingly important. In this regime, the angular coefficients can serve as efficient observables to report experimental measurements and confront theoretical predictions, with the potential to become benchmark legacy measurements of the HL-LHC program.

The remainder of this paper is organized as follows. In~\autoref{sec:tomography}, we outline the quantum tomography framework for the $VV^*$ system and define the spin density matrix and angular observables used throughout the analysis. In~\autoref{sec:entanglement}, we introduce the entanglement measures used in this work. In Sec.~\ref{sec:analysis}, we describe the phenomenological setup and present our results, starting with the leading order analysis and then including higher-order QCD and electroweak corrections. Finally, we summarize our findings and provide an outlook in~\autoref{sec:summary}.

%%%%%%%%%%%%%%%%%%%%%%%%%%%%%%%%%%%%%%%%%%%%%%%%%%%%
\section{Quantum Tomography}
\label{sec:tomography}

The $VV^*$ system arising from Higgs decays $h \to VV^*$ ($V = W, Z$) involves two massive spin--1 particles and admits a natural description, at the level of spin degrees of freedom, within a nine-dimensional Hilbert space. In this representation, the system can be treated as an effective two-qutrit system, and a general spin density matrix can be expanded in terms of normalized irreducible tensor operators $T_M^L$ (with rank $L = 1, 2$ and component $M = -L, \dots, L$) as~\cite{Aguilar-Saavedra:2022wam,Barr:2024djo}
%-----
\begin{equation}
    \rho
    \;=\;
    \frac {1}{9}\left( \mathbb{I}_3 \otimes\mathbb{I}_3
    +A^{1}_{LM} T^{L}_{M}\otimes \mathbb{I}_3
    + A^{2}_{LM}\mathbb{I}_3\otimes T^{L}_{M}
    + C_{L_1 M_1 L_2 M_2} T^{L_1}_{M_1}\otimes T^{L_2}_{M_2}\right),
\label{eq:rho}
\end{equation}
%-----
where repeated indices are summed. The coefficients $A^i_{LM}$ and $C_{L_1 M_1 L_2 M_2}$ encode, respectively, the polarization of each vector boson and their spin–spin correlations. The operators $T^L_M$ provide a complete orthonormal basis satisfying
%-----
\begin{equation}
    (T^L_M)^\dagger = (-1)^M T^L_{-M},
    \qquad
    \Tr\!\left[ T^L_M (T^L_M)^\dagger \right] = 3 .
\end{equation}
%-----

The spin information of the vector bosons is encoded in the angular distributions of their decay products. To access these correlations, we consider semi-leptonic Higgs decay channels, $h \to ZZ^* \to \ell^+\ell^- q\bar q$, with $q=c,b$, and
$h \to WW^* \to \ell^\pm \nu_\ell q\bar q'$, with $q,q'=c,s$.
Throughout, we consider the charged lepton final states to be $\ell^\pm=e^\pm,\mu^\pm$. For a vector boson $V$ decaying into a pair of fermions $f f'$, the decay matrix in the helicity basis is given by~\cite{Boudjema:2009fz}
%-----
\begin{equation}
\resizebox{\textwidth}{!}{$
\Gamma
=
\frac{1}{4}
\begin{pmatrix}
1+\delta_f+(1-3\delta_f)\cos^2\theta - 2\eta_f\cos\theta
&
\frac{1}{\sqrt{2}}\Big[(1-3\delta_f)\sin2\theta - 2\eta_f\sin\theta\Big] e^{i\varphi}
&
(1-3\delta_f)(1-\cos^2\theta) e^{2i\varphi}
\\[0.6em]
\frac{1}{\sqrt{2}}\Big[(1-3\delta_f)\sin2\theta - 2\eta_f\sin\theta\Big] e^{-i\varphi}
&
4\delta_f+2(1-3\delta_f)\,\sin^2\theta
&
-\frac{1}{\sqrt{2}}\Big[(1-3\delta_f)\sin2\theta + 2\eta_f\sin\theta\Big] e^{i\varphi}
\\[0.6em]
(1-3\delta_f)(1-\cos^2\theta) e^{-2i\varphi}
&
-\frac{1}{\sqrt{2}}\Big[(1-3\delta_f)\sin2\theta + 2\eta_f\sin\theta\Big] e^{-i\varphi}
&
1+\delta_f+(1-3\delta_f)\cos^2\theta + 2\eta_f\cos\theta
\end{pmatrix}
$}\,,
\label{eq:Gamma_mat}
\end{equation}
%-----
where $\theta$ and $\varphi$ denote the polar and azimuthal angles of the fermion $f$ in the rest frame of the parent vector boson. 

The two decay parameters, $\eta_f$ and $\delta_f$, determine how the polarization of the parent boson is transferred to the angular distribution of the final state. In general, for a spin-1 boson of invariant mass $m$ decaying into fermions with masses $m_1$ and $m_2$, and left and right chiral couplings $C_L$ and $C_R$, they read
%---
\begin{align}
\eta_f &= 
\frac{
2(C_L^2 - C_R^2)\sqrt{1 + (x_1^2 - x_2^2)^2 - 2(x_1^2 + x_2^2)}
}{
12 C_L C_R x_1 x_2 + (C_R^2 + C_L^2)\left[2 - (x_1^2 - x_2^2)^2 + (x_1^2 + x_2^2)\right]
},
\label{eq:eta}
\\
\delta_f &=
\frac{
4 C_L C_R x_1 x_2 + (C_R^2 + C_L^2)\left[(x_1^2 + x_2^2) - (x_1^2 - x_2^2)^2\right]
}{
12 C_L C_R x_1 x_2 + (C_R^2 + C_L^2)\left[2 - (x_1^2 - x_2^2)^2 + (x_1^2 + x_2^2)\right]
},
\label{eq:delta}
\end{align}
%---
where $x_i=m_i/m$.

In the limit of massless fermions, these expressions simplify to
\begin{equation}
\eta_f \to \frac{C_L^2 - C_R^2}{C_R^2 + C_L^2},
\qquad
\delta_f \to 0,
\end{equation}
so that the spin analyzing power $\eta_f$ is determined solely by the chiral structure of the interaction. For example, consider the $W^-$ decays. The $W^-$ couples only to left handed fermions and right handed antifermions, leading to a strong correlation between the fermion direction and the $W^-$ spin. In particular, the charged lepton $\ell^-$ and the down-type quark $d$ are emitted preferentially along the spin direction, while the corresponding antifermions, $\bar\nu_\ell$ and $\bar u$, are emitted preferentially opposite to it. In the massless limit, this correlation becomes maximal, yielding $\eta_{\ell^-} = \eta_d = +1$ and $\eta_{\bar\nu_\ell} = \eta_{\bar u} = -1$. The pattern is reversed for $W^+$ decays.
 In contrast, for $Z$ boson decays both chiral couplings are nonzero, leading to reduced spin analyzing powers that depend on the fermion types. The explicit expressions for $\eta_f$ 
 for the final states considered in this work are summarized in \autoref{tab:spins_Z}.

%----------
\begin{table}[h!]
    \centering
    \begin{tabular}{c|c|cccc}
    \toprule
    $V$ & $f$ & $C_L$ & $C_R$ & $\eta_f$ \\ 
    \midrule
    \multirow{3}{*}{$Z$} 
        & $\ell^-$ & $-\frac12 + s_W^2$ & $s_W^2$ & $\frac{1-4 s_W^2}{1-4 s_W^2+8 s_W^4}$  \\
        & $c$ & $\frac12 - \frac23 s_W^2$ & $-\frac23 s_W^2$ & $\frac{9-24 s_W^2}{9-24 s_W^2+32 s_W^4}$  \\
        & $b$ & $-\frac12 + \frac13 s_W^2$ & $\frac13 s_W^2$ & $\frac{(9-12 s_W^2)}{9-12 s_W^2+8 s_W^4}$  \\
    \bottomrule
    \end{tabular}
    \caption{Left and right chiral couplings $C_L$ and $C_R$, together with the spin analyzing power $\eta_f$, for fermions $f$ produced in $Z$ decays evaluated in the $Z$ rest frame. We define $s_W^2=\sin^2\theta_W$. We adopt the massless approximation for the final state fermions, consistent with the input parameters used in our analysis. See \autoref{tab:etaNLO} for the numerical results of  $\eta_f$ at LO and NLO electroweak. For $W^-$ decays one has $\eta_{\ell^-}=\eta_d=+1$, while the corresponding antiparticles satisfy $\eta_{\bar\nu_\ell} = \eta_{\bar u} = -1$. The signs are reversed for the $W^+$ decay.}
    \label{tab:spins_Z}
\end{table}
%----------

The polar and azimuthal angles $(\theta,\phi)$ of the leptons and jets are defined in the helicity basis~\cite{Bernreuther:2015yna,Aguilar-Saavedra:2022wam}:
\begin{itemize}
\item  The $\hat{z}$ axis is chosen along the direction of the vector boson $V_1$ in the reconstructed $VV^*$ rest frame, where $V_1$ denotes the leptonically decaying $Z$ boson in the $h\to ZZ^*$ channel or the $W^+$ boson in the $h\to WW^*$ channel. 
\item $\hat{x}=\mathrm{sign}(\cos\theta_{\mathrm{CM}})(\hat{p}-\cos\theta_{\mathrm{CM}}\hat{z})/\sin\theta_{\mathrm{CM}}$, with $\hat{p}=(0,0,1)$ and $\cos\theta_{\mathrm{CM}}=\hat{z}\cdot\hat{p}$.
\item $\hat{y}=\hat{z}\times\hat{x}$.
\end{itemize}

Within this setup, the normalized differential distribution can be factorized in terms of the spin density matrix $\rho$ and the decay matrices $\Gamma_{1,2}$:
%-----
\begin{align}
    \frac{1}{\Gamma} \frac{d\Gamma}{d\Omega_1 d\Omega_2} 
    &= \left( \frac{3}{4\pi} \right)^2 
    \Tr\left\{ \rho \, \big(\Gamma_1 \otimes \Gamma_2 \big)^T \right\} \\
    &= \frac{1}{(4\pi)^2} \Bigg[
        1 + A^1_{LM} B_L Y^M_L(\Omega_1)
          + A^2_{LM}  B_L Y^M_L(\Omega_2) \nonumber\\
    &\hspace{1cm} + C_{L_1 M_1 L_2 M_2} B_{L_1}  B_{L_2} 
          Y^{M_1}_{L_1}(\Omega_1) Y^{M_2}_{L_2}(\Omega_2)
      \Bigg],    
\label{eq:dsigma_semi}
\end{align}
%-----
where $\Gamma_1 \equiv \Gamma(\theta_1,\varphi_1)$ and $\Gamma_2 \equiv \Gamma(\theta_2,\varphi_2)$ are the decay matrices of the vector bosons $V_1$ and $V_2$, respectively. The coefficients $B_L$ are determined by the decay parameters, with $B_1=-\sqrt{2\pi}\,\eta_f$ and $B_2=\sqrt{2\pi/5}\,(1-3\delta_f)$.
The angular coefficients $A^i_{LM}$ and $C_{L_1 M_1 L_2 M_2}$ entering this distribution can be obtained using the orthogonality of spherical harmonics:
%-----
\begin{align}
\frac{1}{\sigma} \int \frac{d\sigma}{d\Omega_1 d\Omega_2} 
Y_L^M(\Omega_i)^* \, d\Omega_1 d\Omega_2 
&= \frac{B_L}{4\pi} A^i_{LM}\equiv f_{LM}^i, \label{eq:coef1}\\
\frac{1}{\sigma} \int \frac{d\sigma}{d\Omega_1 d\Omega_2} 
Y_{L_1}^{M_1}(\Omega_1)^* Y_{L_2}^{M_2}(\Omega_2)^* \, d\Omega_1 d\Omega_2 
&= \frac{B_{L_1} B_{L_2}}{(4\pi)^2} C_{L_1 M_1 L_2 M_2}\equiv g_{L_1 M_1 L_2 M_2}.
\label{eq:coef2}
\end{align}
%-----
These relations also define the raw angular moments $f$ and $g$, which can be directly extracted from the angular distribution. They provide a complete characterization of the angular structure for $L \le 2$, without relying on the two-qutrit interpretation.

Within the QT framework, the measured moments can be mapped to the production multipoles $A$ and $C$ of the two-qutrit density matrix in Eq.~\eqref{eq:rho}. This mapping depends on the decay parameters $\eta_f$ and $\delta_f$ and assumes that the intermediate bosons behave as pure spin-1 states.
In the on-shell regime, finite fermion mass effects preserve this spin-1 structure and can be incorporated through mass-corrected expressions for $\eta_f$ and $\delta_f$, so that the mapping between $(f,g)$ and $(A,C)$ amounts to a rescaling, satisfying \Cref{eq:coef1,eq:coef2}. 
 The situation changes when the vector boson becomes off–shell and the fermion mass is nonzero. In this case, the scalar polarization component of the virtual vector boson can couple to the fermion current~\cite{Peskin:1995ev}, generating pure scalar and scalar–vector interference terms contributions. While these contributions populate only $L=0$ and $L=1$ moments, they correspond to additional structures not contained in a pure two-qutrit density matrix in~\Cref{eq:rho}. Hence, interpreting the measured moments solely in terms of the multipoles $A$ and $C$ of a two-qutrit density matrix is no longer exact. Presenting the results in terms of $f$ and $g$ therefore isolates the observable angular information, whereas any mapping to $A$ and $C$ relies on the assumption that scalar contributions are negligible. 

For $m_h=125~\text{GeV}$~\cite{ATLAS:2023oaq}, at least one of the two vector bosons in $h \to VV^*$ is necessarily off-shell. In the semi-leptonic channels considered here, the leptonic decay involves first and second generation fermions, which can be effectively treated as massless. Since the scalar component couples proportionally to the fermion mass, its contribution vanishes in this limit. In the hadronic channel, however, the massless approximation is not always equally accurate, for instance in $h\to ZZ^* \to \ell^+\ell^- b\bar b$. Nevertheless, by restricting the analysis to the near on-shell region of the hadronic vector boson, the scalar contribution remains strongly suppressed. In this regime, both vector bosons can be consistently treated as spin-1 states, allowing the $VV^*$ system to be described as an effective two-qutrit state. A detailed quantitative analysis of these mass effects is presented in \autoref{subsec:LOdecay} (and \autoref{app:spin0}).

%%%%%%%%%%%%%%%%%%%%%%%%%%%%%%%%%%%%%%%%%%%%%%%%%%%%
\section{Quantum Entanglement}
\label{sec:entanglement}

Once the spin density matrix $\rho$ of the $VV^*$ system has been reconstructed through quantum tomography, it can be used to probe genuine quantum correlations between the two vector bosons, in particular quantum entanglement. As a quantitative measure of entanglement, we employ the concurrence $\mathcal{C}(\rho)$~\cite{PhysRevA.64.042315,PhysRevLett.80.2245,PhysRevA.54.3824}.

For generic mixed states of two spin-1 particles, corresponding to two-qutrit systems, the concurrence cannot be expressed in a closed analytic form. Nevertheless, its value can be constrained using lower and upper bounds, denoted by $\mathscr{C}_{\mathrm{LB}}$ and $\mathscr{C}_{\mathrm{UB}}$~\cite{PhysRevLett.98.140505,PhysRevA.78.042308,Fabbrichesi:2023cev}:
%-----
%-----
\begin{align}
    \left(\mathcal{C}(\rho)\right)^2
    &\geq 
    2\, \max\Big\{ 0,\, 
    \Tr[\rho^2] - \Tr[\rho_A^2],\, 
    \Tr[\rho^2] - \Tr[\rho_B^2] \Big\} 
    \;\equiv\; 
    \mathscr{C}^{2}_{\mathrm{LB}}, \nonumber\\
    \left(\mathcal{C}(\rho)\right)^2
    &\leq 
    2\, \min\Big\{ 
    1 - \Tr[\rho_A^2],\, 
    1 - \Tr[\rho_B^2] 
    \Big\}
    \;\equiv\; 
    \mathscr{C}^{2}_{\mathrm{UB}},
\end{align}
%-----
where $\rho_A=\Tr_B(\rho)$ and $\rho_B = \Tr_A(\rho)$ are the reduced density matrices of the two subsystems. A nonvanishing lower bound, $\mathscr{C}_{\mathrm{LB}}>0$, provides evidence that the spin state of one vector boson cannot be described independently of the other, signaling genuine quantum correlations in the Higgs decay. In contrast,  if the upper bound satisfies $\mathscr{C}_{\mathrm{UB}}=0$, the state is fully separable. Therefore, these bounds provide a practical proxy for quantifying the entanglement in the $VV^*$ system.

%%%%%%%%%%%%%%%%%%%%%%%%%%%%%%%%%%%%%%%%%%%%%%%%%%%%
\section{Analysis}
\label{sec:analysis}

\begin{figure}[!tbp]
    \centering
   \includegraphics[width=0.7\textwidth]{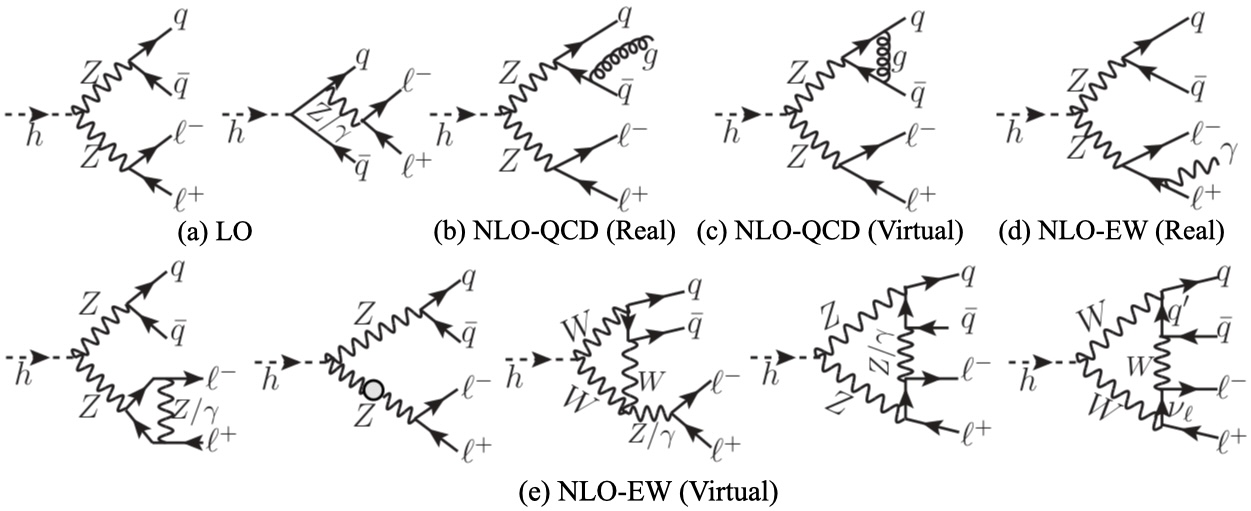}
    \caption{Representative Feynman diagrams for the process $h \to \ell^+\ell^- q\bar{q}$ at LO (a), NLO QCD (b-c), and NLO electroweak (d-e).}
    \label{fig:Feyn_h_2j_2j}
\end{figure}

The quantum numbers of the Higgs boson impose strong constraints on the structure of the diboson spin density matrix for the $h\to VV^*$ channel, ensuring the presence of spin entanglement at leading order (LO) precision even when integrating over the full phase space~\cite{Barr:2021zcp,Aguilar-Saavedra:2022wam,Fabbrichesi:2023cev,Fabbrichesi:2023jep,Fabbri:2023ncz,Bernal:2023ruk,Aguilar-Saavedra:2024whi,Subba:2024mnl,Bernal:2024xhm,Sullivan:2024wzl,Aguilar-Saavedra:2024jkj}. Incorporating next-to-leading order (NLO) effects modifies this structure in a channel dependent way. In particular, the $h\to e^+e^-\mu^+\mu^-$ decay shows distortions that challenge the validity of the two-qutrit description~\cite{Grossi:2024jae,DelGratta:2025qyp,Goncalves:2025mvl,DelGratta:2025xjp,Aguilar-Saavedra:2025byk}, whereas the $h \to e^+ \nu_e \mu^- \bar{\nu}_{\mu}$ channel remains comparatively stable under higher-order corrections, granting a more robust interpretation of quantum observables, such as entanglement, using a two-qutrit system~\cite{Goncalves:2025xer}. This distinct channel dependence provides further motivation to investigate a broader set of final states. In this section, we extend the analysis of the fully leptonic Higgs decays to the semi-leptonic channels, $h\to \ell^\pm \nu_\ell q\bar{q}'$ and $h\to \ell^+\ell^- q\bar{q}$, and study the impact of final state fermion masses as well as higher-order QCD and electroweak (EW) corrections. Representative Feynman diagrams for these processes are shown in \Cref{fig:Feyn_h_lvjj,fig:Feyn_h_2j_2j}.

%----------
\begin{figure}[!tbp]
    \centering
   \includegraphics[width=0.7\textwidth]{Figures/h_WW_Feynman.jpeg}
    \caption{Representative Feynman diagrams for the process $h\to \ell^-\bar{\nu}_\ell q \bar{q}^\prime$ at LO (a), NLO QCD (b-c), and NLO electroweak (d-e).}
    \label{fig:Feyn_h_lvjj}
\end{figure}
%----------

%----------
%----------

%%%%%%%%%%%%
\subsection{Leading Order Decay}
\label{subsec:LOdecay}

We consider the semi-leptonic decay modes from the Higgs boson $h \to  \ell^+\ell^- q\bar q$, with $q=c,b$, and $h \to \ell^\pm \nu_\ell q\bar q'$, with $q,q'=c,s$. The charged leptons in the final state are taken to be $\ell = e,\mu$.  Events are generated with \texttt{MadGraph5\_aMC@NLO}~\cite{Alwall:2014hca,Frederix:2018nkq}, using the \texttt{loop\_qcd\_qed\_sm\_Gmu} UFO model. Unstable particles are treated within the complex mass scheme~\cite{Denner:1999gp,Denner:2005fg}, and the following on-shell parameters are used 
%-----
\begin{align}
    m_W &=  80.385~\mathrm{GeV}, && \Gamma_W = 2.085~\mathrm{GeV}, \label{eq:CMparams1}\\
    m_Z &=  91.1876~\mathrm{GeV}, && \Gamma_Z =  2.4952~\mathrm{GeV}, \label{eq:CMparams2}\\
    G_\mu &= 1.1663787\times10^{-5}~\mathrm{GeV}^{-2}, && \alpha_s(M_Z)=0.118\,, \label{eq:GFparam}    \\
    m_h &= 125~\mathrm{GeV}, && \Gamma_h = 4.097~\mathrm{MeV}, \label{eq:mh}\\
    m_t &= 173.2~\mathrm{GeV}, && \Gamma_t =  1.369~\mathrm{GeV}. \label{eq:mt} 
\end{align}
%-----
The renormalization scale is set to $\mu_R=m_h$. All fermions, except the top quark, are considered massless in our simulation. In the study of final state fermion mass effects presented in \autoref{subsec:LOdecay} (and~\autoref{app:spin0}), we also show results obtained by varying the $c$ and $b$ quark masses, as explicitly indicated. The above on-shell parameters for unstable particles are converted into the pole values used as input to the complex mass scheme with 
%----
\begin{align}
M = \frac{M^{\mathrm{OS}}}{\sqrt{1+\left(\Gamma^\mathrm{OS}/M^\mathrm{OS}\right)^2}}\,, && \Gamma = \frac{\Gamma^\mathrm{OS}}{\sqrt{1+\left(\Gamma^\mathrm{OS}/M^\mathrm{OS}\right)^2}}\,.
\end{align}
%----

Since the QT procedure requires specifying the decay parameters $\eta_f$ and $\delta_f$, which differ for up and down-type quarks, see~\Cref{eq:eta,eq:delta}, we treat the $\ell^+\ell^- q\bar q$ sample separately for the $b\bar b$ and $c\bar c$ final states.
The choice of spin analyzers is then fixed by the decay channel. In the $\ell^\pm\nu_\ell q\bar q'$ channel, we use the charged lepton together with the down-type quark, namely the $s$ or $\bar{s}$ quark. In the $\ell^+\ell^- q\bar q$ channel, we instead select the negatively charged lepton and the $b$ or $c$ quark as analyzers.\footnote{Rather than focusing on the technical aspects of neutrino reconstruction~\cite{Fabbri:2023ncz,Cho:2008tj,Barr:2010zj,CMS:2017znf,Goncalves:2018agy,ATLAS:2022ooq,CMS:2022uhn,Ackerschott:2023nax,CMS:2024bua,Dong:2024fzt,ATLAS:2025abg}, jet flavor tagging~\cite{Qu:2022mxj,Fabbri:2023ncz,ATLAS:2023lwk,CMS-PAS-BTV-22-001}, or charge tagging~\cite{ATLAS:2015rlw,CMS:2017yer,Gianelle:2022unu,Kang:2023ptt,Dong:2023xiw,Dong:2024xsb} (see also~\autoref{app:chargeeffect}), whose performance depends on the reconstruction strategy and detector response, we assume ideal knowledge of the final state momenta throughout this study. This idealized setup allows us to isolate the impact of mass effects, and NLO QCD and electroweak corrections on the angular coefficients. Thus, the resulting corrected coefficients obtained here define a theoretical benchmark that realistic reconstruction strategies should aim to approximate.} 

%----------
\begin{table}[tb!]
\centering
\begin{tabular}{l|ccc}
\toprule
\multicolumn{4}{c}{$h\to \ell^+\nu_\ell \bar{c}s$} \\
\midrule
 & $m_c=0$ & $m_c=1.3$ GeV & Ratio \\
\midrule
\multicolumn{4}{c}{Inclusive} \\
\midrule
$f_{2,0}^1$ & $-0.0517(1)$ & $-0.0513(1)$ & $0.992(3)$ \\
$f_{2,0}^2$ & $-0.0516(1)$ & $-0.0506(1)$ & $0.980(3)$ \\
$g_{1,0,1,0}$ & $0.02364(4)$ & $0.02369(4)$ & $1.002(2)$ \\
$g_{1,1,1,-1}$ & $-0.03772(3)$ & $-0.03766(3)$ & $0.998(1)$ \\
$g_{2,1,2,-1}$ & $-0.00753(3)$ & $-0.00749(3)$ & $0.994(5)$ \\
$g_{2,0,2,0}$  & $0.01128(3)$ & $0.01116(3)$ & $0.989(3)$ \\
$g_{2,2,2,-2}$ & $0.00471(3)$ & $0.00470(3)$ & $0.997(8)$ \\ 
\midrule
\multicolumn{4}{c}{On-shell hadronic selection: $|m_{q\bar{q}'}-m_W|<10~\text{GeV}$} \\
\midrule
$f_{2,0}^1$ & $-0.0503(1)$ & $-0.0499(1)$ & $0.992(3)$ \\
$f_{2,0}^2$ & $-0.0503(1)$ & $-0.0499(1)$ & $0.992(3)$ \\
$g_{1,0,1,0}$ & $0.02401(4)$ & $0.02411(4)$ & $1.004(2)$ \\
$g_{1,1,1,-1}$ & $-0.03803(3)$ & $-0.03800(3)$ & $0.999(1)$ \\
$g_{2,1,2,-1}$ & $-0.00757(3)$ & $-0.00761(3)$ & $1.005(5)$ \\
$g_{2,0,2,0}$  & $0.01111(3)$ & $0.01114(3)$ & $1.003(3)$ \\
$g_{2,2,2,-2}$ & $0.00484(3)$ & $0.00483(3)$ & $0.997(8)$ \\
\bottomrule
\end{tabular}
\caption{Angular moments $f$ and $g$ defined in \Cref{eq:coef1,eq:coef2} for $h\to \ell^+\nu_\ell \bar{c}s$ at LO, comparing the massless approximation to simulations with finite charm quark mass. Results are shown for an inclusive selection (upper block) and for an on-shell hadronic selection (lower block). Statistical uncertainties are shown in parentheses.}
\label{tab:mass_WW_combined}
\end{table}
%----------

%%%%%%%%%%%
\subsubsection{Quark Mass Effects on Angular Observables}
\label{sec:mbEffects}
We first study the impact of finite quark masses on the angular coefficients. Rather than presenting the extracted QT coefficients $A$ and $C$, we show the raw angular moments $f$ and $g$, see \Cref{eq:coef1,eq:coef2}. This choice allows us to directly probe the observable angular structure, independent of the two-qutrit interpretation. The mapping from $(f,g)$ to $(A,C)$ relies on the assumption that scalar contributions are negligible and that the two-qutrit description is satisfied, see the discussion in \autoref{sec:tomography} following \autoref{eq:coef2}. After establishing the validity of this mapping in the present section for a particular kinematic regime, we present results for $A$ and $C$ in \Cref{sec:LO_entanglement}.

 %--------
\begin{table}[tb!]
\centering
\resizebox{\textwidth}{!}{
\begin{tabular}{l|ccc|ccc}
\toprule
& \multicolumn{3}{c|}{$h\to \ell^+\ell^- b\bar b$}
& \multicolumn{3}{c}{$h\to \ell^+\ell^- c\bar c$} \\
\cmidrule(lr){2-4} \cmidrule(lr){5-7}
& $m_b=0$ & $m_b=4.5$ GeV & Ratio 
& $m_c=0$ & $m_c=1.3$ GeV & Ratio \\
\midrule
\multicolumn{7}{c}{Selections: $m_{\ell\ell}>10~\text{GeV}$, $m_{q\bar q}>10~\text{GeV}$} \\
\midrule
$f_{2,0}^1$ & $-0.04794(9)$ & $-0.04839(9)$ & $1.009(2)$ & $-0.04801(9)$ & $-0.04795(9)$ & $0.999(3)$ \\
$f_{2,0}^2$ & $-0.04798(9)$ & $-0.04093(9)$ & $0.853(2)$ & $-0.04785(9)$ & $-0.04706(9)$ & $0.983(3)$ \\
$g_{1,0,1,0}$ & $-0.00498(2)$ & $-0.00471(2)$ & $0.946(7)$ & $-0.00367(2)$ & $-0.00374(2)$ & $1.019(8)$ \\
$g_{1,1,1,-1}$ & $0.00781(2)$ & $0.00747(2)$ & $0.956(3)$ & $0.00582(2)$ & $0.00585(2)$ & $1.005(5)$ \\
$g_{2,1,2,-1}$ & $-0.00776(2)$ & $-0.00703(2)$ & $0.906(3)$  & $-0.00775(2)$ & $-0.00766(2)$ & $0.988(3)$ \\
$g_{2,0,2,0}$  & $0.01094(2)$ & $0.00978(2)$ & $0.893(2)$ & $0.01097(2)$ & $0.01081(2)$ & $0.985(3)$ \\
$g_{2,2,2,-2}$ & $0.00492(2)$ & $0.00454(2)$ & $0.922(5)$ & $0.00487(2)$ & $0.00484(2)$ & $0.993(5)$ \\ 
\midrule
\multicolumn{7}{c}{Selections:  $m_{\ell\ell}>20~\text{GeV}$, $|m_{q\bar q}-m_Z|<10~\text{GeV}$} \\
\midrule
$f_{2,0}^1$ & $-0.02915(9)$ & $-0.02936(9)$ & $1.007(4)$ & $-0.0294(1)$ & $-0.0293(1)$ & $0.997(4)$ \\
$f_{2,0}^2$ & $-0.02925(9)$ & $-0.02911(9)$ & $0.995(4)$ & $-0.0292(1)$ & $-0.0293(1)$ & $1.003(4)$ \\
$g_{1,0,1,0}$ & $-0.00622(3)$ & $-0.00609(3)$ & $0.979(7)$ & $-0.00454(3)$ & $-0.00459(3)$ & $1.011(9)$ \\
$g_{1,1,1,-1}$ & $0.00839(2)$ & $0.00835(2)$ & $0.995(3)$ & $0.00622(2)$ & $0.00622(2)$ & $1.000(4)$ \\
$g_{2,1,2,-1}$ & $-0.00833(2)$ & $-0.00819(2)$ & $0.983(3)$ & $-0.00828(2)$ & $-0.00826(2)$ & $0.997(3)$ \\
$g_{2,0,2,0}$  & $0.00977(2)$ & $0.00964(3)$ & $0.986(4)$ & $0.00980(3)$ & $0.00972(3)$ & $0.991(4)$ \\
$g_{2,2,2,-2}$ & $0.00612(2)$ & $0.00604(2)$ & $0.987(4)$ & $0.00607(2)$ & $0.00607(2)$ & $1.000(5)$ \\ 
\bottomrule
\end{tabular}
}
\caption{Angular moments defined in \Cref{eq:coef1,eq:coef2} for $h\to \ell^+\ell^- q\bar q$ at LO, comparing the massless approximation to simulations with finite quark masses ($m_b=4.5~\text{GeV}$, $m_c=1.3~\text{GeV}$). Results are shown for a more inclusive selection (upper block) and for an on-shell hadronic selection (lower block).  Statistical uncertainties are shown in parentheses.}
\label{tab:mass_ZZ_combined}
\end{table}
%---------

We begin with the $h\to WW^*$ channel, where the leading mass effects arise from charm quarks in the hadronic decay $W^-\to \bar c s$. As shown in \autoref{tab:mass_WW_combined}, including a finite charm quark mass, $m_c=1.3~\text{GeV}$, modifies the angular moments at the $\mathcal{O}(2\%)$ level or below in the inclusive setup, where no constraint is imposed on the vector boson invariant masses. Requiring the hadronic $W$ decay to lie near the on-shell region, $|m_{cs}-m_W|<10~\text{GeV}$, further suppresses these effects to the sub-percent level. 
In the following analysis, we therefore adopt this kinematic selection, for which the massless quark approximation provides an excellent description of the $h\to WW^*$ channel.\footnote{In \autoref{sec:NLOEW}, we show that the near on-shell regime of one of the gauge bosons also plays an important role in suppressing higher-order electroweak contributions that can potentially challenge the two-qutrit description~\cite{Goncalves:2025mvl,Goncalves:2025xer}.}

We now turn to the $h\to ZZ^*$ channel. We first consider more inclusive selections, imposing $m_{\ell\ell}>10~\text{GeV}$ and $m_{q\bar q}>10~\text{GeV}$. In this setup, 
the hadronic invariant mass $m_{q\bar q}$ can lie well below the $m_Z$, so that the ratio $m_b/m_{b\bar b}$ (or $m_c/m_{c\bar c}$) is not parametrically suppressed by the $Z$ boson mass. As shown in \autoref{tab:mass_ZZ_combined}, several moments shift by $\mathcal{O}(5\text{--}15\%)$ when comparing $m_b=0$ and $m_b=4.5~\text{GeV}$. Charm-mass effects remain more modest, below about $\mathcal{O}(4\%)$.
Similarly to the $h\to WW^*$ channel, these distortions are predominantly driven by configurations with small $m_{q\bar q}$ and can be mitigated by selecting the on-shell regime for the hadronic decay, $|m_{q\bar q}-m_Z|<10~\text{GeV}$. After this selection, quark mass effects are suppressed to below $\mathcal{O}(2\%)$ for $b$-quarks, and to the sub-percent level for $c$-quarks, as shown in the lower part of \autoref{tab:mass_ZZ_combined}.

%---------------
\begin{table}[t]
\centering
\small
\setlength{\tabcolsep}{1pt}
\resizebox{\textwidth}{!}{
\begin{tabular}{l|ccc|ccc|ccc}
\toprule
 & \multicolumn{3}{c|}{$h\to \ell^+\nu_\ell \bar{c} s$}
 & \multicolumn{3}{c|}{$h\to \ell^+\ell^- c\bar{c}$}
 & \multicolumn{3}{c}{$h\to \ell^+\ell^- b\bar{b}$} \\
\cmidrule(lr){2-4}\cmidrule(lr){5-7}\cmidrule(lr){8-10}
 & $m_c=0$ & $m_c=1.3$~GeV & Ratio
 & $m_c=0$ & $m_c=1.3$~GeV & Ratio
 & $m_b=0$ & $m_b=4.5$~GeV & Ratio \\
\midrule
$A_{2,0}^1$ & $-0.564(1)$ & $-0.560(1)$ & $0.993(2)$
            & $-0.329(1)$ & $-0.329(1)$ & $1.000(4)$
            & $-0.327(1)$ & $-0.329(1)$ & $1.006(4)$ \\

$A_{2,0}^2$ & $-0.564(1)$ & $-0.560(1)$ & $0.993(2)$
            & $-0.328(1)$ & $-0.329(1)$ & $1.003(4)$
            & $-0.327(1)$ & $-0.328(1)$ & $1.003(4)$ \\

$C_{1,0,1,0}$ & $-0.6035(9)$ & $-0.6062(9)$ & $1.004(4)$
              & $-0.766(4)$ & $-0.773(4)$ & $1.009(7)$
              & $-0.774(3)$ & $-0.764(3)$ & $0.987(5)$ \\

$C_{1,1,1,-1}$ & $0.9560(9)$ & $0.9555(9)$ & $0.999(2)$
               & $1.047(4)$ & $1.048(4)$ & $1.001(5)$
               & $1.046(3)$ & $1.046(3)$ & $1.000(4)$ \\

$C_{2,1,2,-1}$ & $-0.952(3)$ & $-0.956(3)$ & $1.004(4)$
               & $-1.041(3)$ & $-1.039(3)$ & $0.998(4)$
               & $-1.046(3)$ & $-1.035(3)$ & $0.989(4)$ \\

$C_{2,0,2,0}$  & $1.396(4)$ & $1.400(4)$ & $1.003(4)$
               & $1.233(3)$ & $1.223(3)$ & $0.992(3)$
               & $1.228(4)$ & $1.218(4)$ & $0.991(4)$ \\

$C_{2,2,2,-2}$ & $0.607(4)$ & $0.607(4)$ & $1.000(9)$
               & $0.764(3)$ & $0.764(3)$ & $1.000(5)$
               & $0.770(3)$ & $0.763(3)$ & $0.991(5)$ \\

\bottomrule
\end{tabular}
}
\caption{Angular coefficients at LO for the channels $h\to \ell^+\nu_\ell \bar{c}s$, $h\to \ell^+\ell^- c\bar c$, and $h\to \ell^+\ell^- b\bar b$. For all channels, we impose the on-shell hadronic selection $|m_{q\bar{q}^{(\prime)}}-m_V|<10$~GeV. For the $h\to \ell^+\ell^- q\bar q$ modes, we additionally require $m_{\ell\ell}>20$~GeV. The coefficients are presented in the massless approximation and including finite quark masses. 
In the expressions for the decay parameters $\delta_q$ and $\eta_q$, we approximate $m_{q\bar q} = m_V$ for $x_q \neq 0$, so that $x_q = m_q/m_V$.}
\label{tab:mass_A_C_coeffs}
\end{table}
%---------------

In addition to the scalar component in hadronic decays, massive fermions can generate extra  contributions at leading order that may spoil the two-qutrit description in the $h\to \ell^+\ell^- q\bar q$ channel. This decay receives contributions from diagrams where charm or bottom quarks are produced directly through their Yukawa coupling, while the leptons arise from $Z/\gamma$ radiation off the quark lines, as illustrated on the right side of \autoref{fig:Feyn_h_2j_2j}~(a). The kinematic selections adopted in our analysis are designed to suppress these configurations. Specifically, the requirements $m_{\ell\ell}>20~\text{GeV}$ and $|m_{q\bar{q}}-m_Z|<10~\text{GeV}$ deplete these contributions, consistent with the small mass effects observed in \autoref{tab:mass_ZZ_combined}.

%%%%%%%%%%%
\subsubsection{Two-Qutrit Representation and Entanglement}
\label{sec:LO_entanglement}

After establishing that the heavy quark mass effects can be suppressed to the percent level or below by imposing suitable kinematic selections, we now examine whether the selected events admit an effective two-qutrit description. In \autoref{tab:mass_A_C_coeffs}, we present the angular coefficients $A$ and $C$ for the considered channels. The results show that, after the considered selections, the mass effects remain below the $2\%$ level, supporting the validity of the two-qutrit description.
The table also enables a direct comparison across channels.
The two $h \to ZZ^*$ modes at LO, $\ell^+\ell^- b\bar b$ and $\ell^+\ell^- c\bar c$, show only small differences, arising from the distinct $b$ and $c$ quark masses. In contrast, the coefficients in the $h \to ZZ^*$ and $h \to WW^*$ channels can differ by tens of percent, reflecting the different $W$ and $Z$-boson masses and the resulting differences in off-shell kinematics and polarization fractions.

If the system is well described by a two-qutrit density matrix, the angular coefficients must satisfy the symmetry relations imposed by angular momentum conservation and parity invariance in $h\to VV^*$~\cite{Aguilar-Saavedra:2022wam,DelGratta:2025qyp}:
%--------------------------------------
\begin{align}
    A^{1}_{20} &= A^{2}_{20} \neq 0\,,\nonumber\\
    C_{1,1,1,-1} &= C_{1,-1,1,1} = -C_{2,1,2,-1} = -C_{2,-1,2,1} \neq 0\,,\label{eq:LOConds}\\
    C_{2,2,2,-2} &= C_{2,-2,2,2} = -C_{1,0,1,0} = 2 - C_{2,0,2,0} \neq 0\,,\nonumber\\
    \frac{A^{1,2}_{20}}{\sqrt{2}} + 1 &= C_{2,2,2,-2}\,.\nonumber
\end{align}
%--------------------------------------
These conditions constrain the LO density matrix to the following form:
%--------------------------------------
\begin{align}
    \label{eq:rhoLOForm}
    \rho_{\mathrm{LO}} = \begin{pmatrix}
        0 & 0 & 0 & 0 & 0 & 0 & 0 & 0 & 0 \\
        0 & 0 & 0 & 0 & 0 & 0 & 0 & 0 & 0 \\
        0 & 0 & \frac{1}{3}C_{2,2,2,-2} & 0 & \frac{1}{3}C_{2,1,2,-1} & 0 & \frac{1}{3}C_{2,2,2,-2} & 0 & 0 \\
        0 & 0 & 0 & 0 & 0 & 0 & 0 & 0 & 0 \\
        0 & 0 & \frac{1}{3}C_{2,1,2,-1} & 0 & \frac{1}{3}(3 - 2C_{2,2,2,-2}) & 0 & \frac{1}{3}C_{2,1,2,-1} & 0 & 0 \\
        0 & 0 & 0 & 0 & 0 & 0 & 0 & 0 & 0 \\
        0 & 0 & \frac{1}{3}C_{2,2,2,-2} & 0 & \frac{1}{3}C_{2,1,2,-1} & 0 & \frac{1}{3}C_{2,2,2,-2} & 0 & 0 \\
        0 & 0 & 0 & 0 & 0 & 0 & 0 & 0 & 0 \\
        0 & 0 & 0 & 0 & 0 & 0 & 0 & 0 & 0 \\
    \end{pmatrix},
\end{align}
%--------------------------------------
where only nine entries are nonzero, fully determined by two independent parameters, chosen here as $C_{2,2,2,-2}$ and $C_{2,1,2,-1}$.
The coefficients reported in \autoref{tab:mass_A_C_coeffs} satisfy the relations in \autoref{eq:LOConds} to good accuracy, confirming that, after the kinematic selections used to suppress scalar and other non-two-qutrit contributions (\autoref{fig:Feyn_h_2j_2j}~(a), right diagram), the semi-leptonic channels remain well described by an effective two-qutrit system at LO.\footnote{In \autoref{app:spin0}, we highlight the importance of requiring the hadronically decaying vector boson to be near on-shell by considering the complementary configuration in which the leptonically decaying vector boson is instead selected near its mass shell. In this case, the LO relations in \autoref{eq:LOConds} are violated in the $h\to \ell^+\ell^- b\bar{b}$ channel due to fermion mass contributions that lie beyond the two-qutrit description.}

\begin{figure}[!tb]
    \centering
    \includegraphics[width=0.325\textwidth]{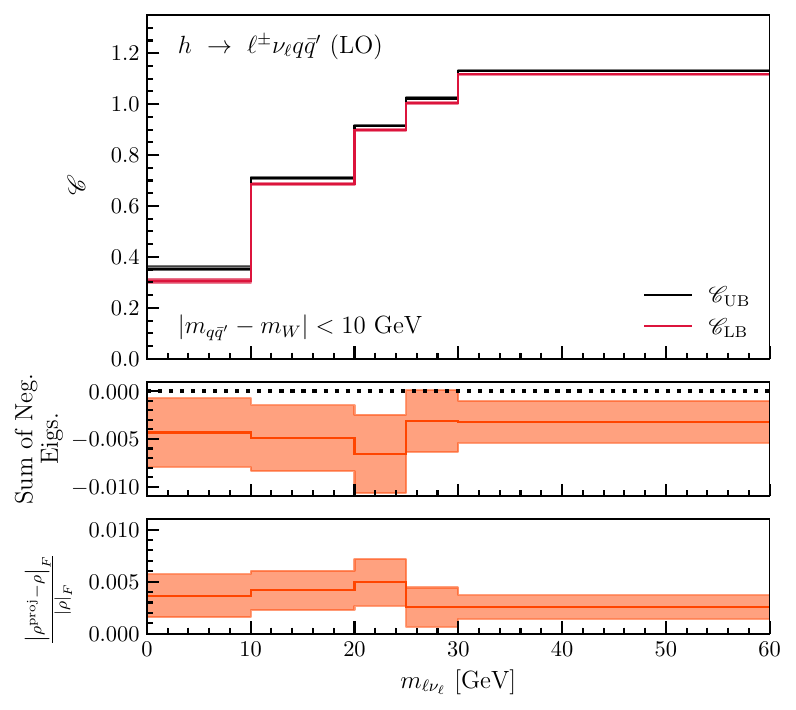}
    \includegraphics[width=0.325\textwidth]{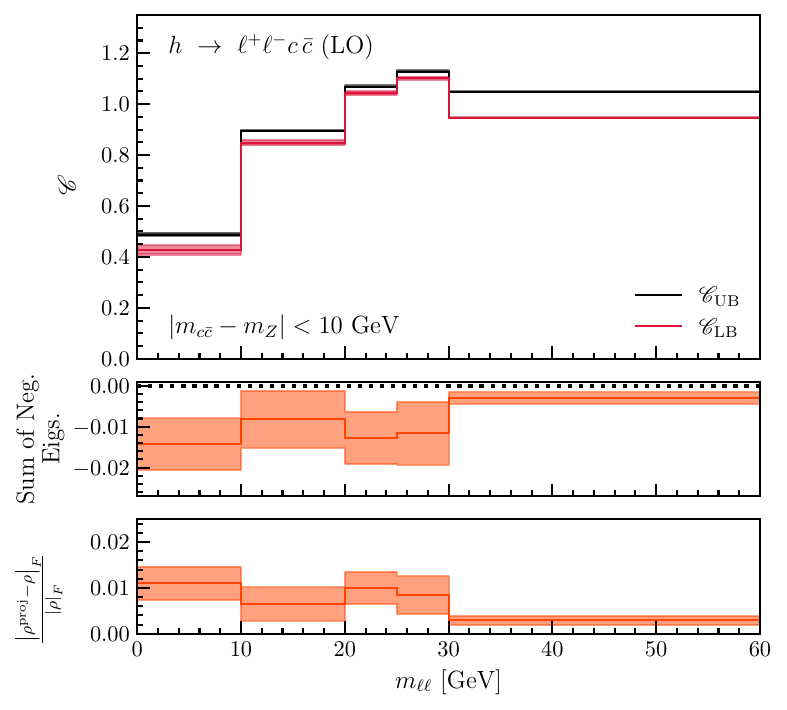} 
    \includegraphics[width=0.325\textwidth]{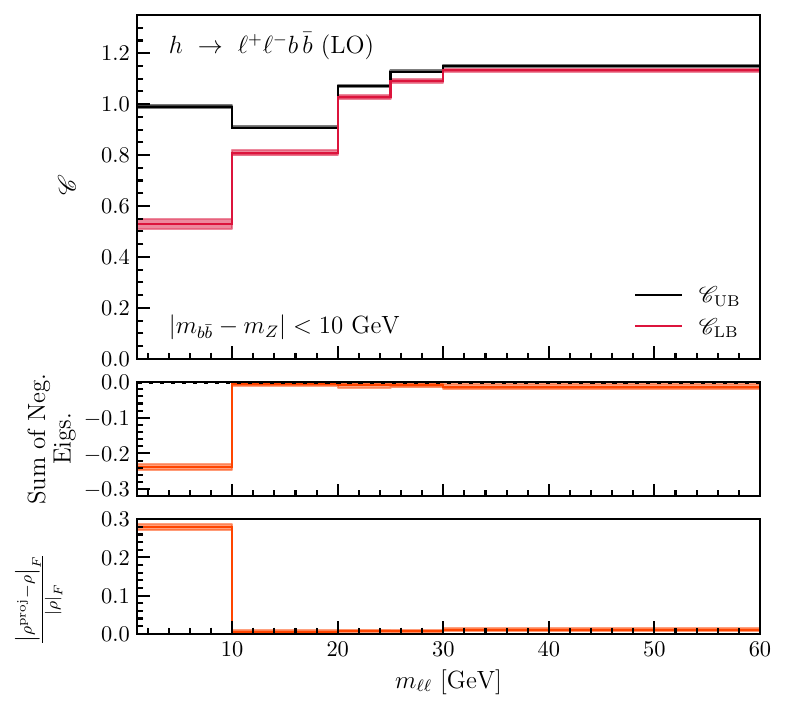} 
    \caption{Upper ($\mathcal{C}_{\rm UB}$, black) and lower ($\mathcal{C}_{\rm LB}$, red) bounds on the concurrence as a function of the invariant mass of the dilepton system. The LO analysis considers the channels $h\to \ell^\pm \nu_\ell q\bar{q}'$ (left panel), $h\to \ell^+\ell^-c\bar{c}$ (central panel), and $h\to \ell^+\ell^-b\bar{b}$ (right panel), with the hadronic invariant mass selected near the on-shell region,  $|m_{q\bar{q}^{(\prime)}}-m_V|<10$~GeV. The input parameters are given in \Cref{eq:CMparams1,eq:CMparams2,eq:GFparam,eq:mh,eq:mt} and $m_b=4.5~\mathrm{GeV}$.
    The middle subpanels of each figure display the sum of the negative eigenvalues for the reconstructed density matrix and the bottom subpanels show the relative distance between the reconstructed density matrix, $\rho$, and the closest positive semi-definite, Hermitian, and unit-trace matrix, $\rho^\text{proj}$, evaluated using the Frobenius norm. The shaded regions represent the statistical uncertainty.}
    \label{fig:CLB-CUB-LO}
\end{figure}

Finally, we quantify the entanglement of the effective two-qutrit state describing the diboson system. The lower and upper bounds of the concurrence are computed for each decay channel following~\autoref{sec:entanglement}. In \autoref{fig:CLB-CUB-LO}, we display these bounds as functions of the invariant mass of dilepton system. For all semi-leptonic Higgs decay channels, both bounds remain strictly positive, confirming the presence of entanglement at leading order precision. The two $h\to ZZ^*$ final states yield nearly identical results, while small differences appear between the $h\to ZZ^*$ and $h\to WW^*$ channels, consistent with the variations in the angular coefficients reported in~\autoref{tab:mass_A_C_coeffs}.

%---
%---

We also probe the stability of the reconstructed density matrix in a differential way as a function of the invariant mass of the dilepton in \autoref{fig:CLB-CUB-LO}. As a diagnostic of physical consistency, we use the sum of negative eigenvalues as a proxy for testing whether the reconstructed density matrix is positive semi-definite, as required for any physical state~\cite{Goncalves:2025mvl,Goncalves:2025xer}. We find that this quantity is very small and consistent with zero within $2\sigma$.\footnote{We propagate the uncertainties in the reconstructed density matrix to the eigenvalues, concurrence bounds, and projected density matrix $\rho^\text{proj}$, using the publicly available Python package \texttt{pyerrors}~\cite{Joswig:2022qfe}, which is based on the $\Gamma$-method~\cite{Wolff:2003sm, Ramos:2018vgu}.} The only exception occurs in the low invariant-mass region, $1~\text{GeV}<m_{\ell\ell}<10$~GeV, for the $h\to \ell^+\ell^- b\bar{b}$ decay. Although this kinematic region is not included in the main analysis, we present it here to illustrate the breakdown of the two-qutrit description and the importance of the invariant mass selection, in particular the requirement $m_{\ell\ell}>20$~GeV adopted, which suppresses the non-two-qutrit contributions at LO discussed above.

In addition to the reconstructed density matrix, $\rho$, obtained from the quantum tomography procedure described in \autoref{sec:tomography}, we define the projected density matrix, $\rho^\text{proj}$, as the closest physical density matrix to $\rho$, namely positive semi-definite, Hermitian, and unit-trace. Comparing $\rho$ with $\rho^\text{proj}$ provides a complementary way to assess the stability of the reconstructed density matrix.
To quantify the distance between the two matrices, we use the Frobenius norm~\cite{10.5555/248979}
%----
\begin{equation}
\left| \rho^{\rm{proj}} - \rho \right|_{F} \equiv \sqrt{\sum\limits_{i,j}\left| \rho^{\rm{proj}}_{ij} - \rho_{ij} \right|^2}.
\label{eq:Frobenius}
\end{equation}
%----
The projection is performed using the QuTiP package~\cite{Johansson:2012qtx}, following the algorithm of Ref.~\cite{Smolin:2011tob}. The bottom subpanels of \autoref{fig:CLB-CUB-LO} display the relative distance, $|\rho^\text{proj}-\rho|_F/|\rho|_F$. Consistent with the behavior of the negative eigenvalues, this quantity remains compatible with zero across the full kinematic range for all channels, with the exception of the first bin in the $h\to \ell^+\ell^- b\bar{b}$ decay, which is excluded in our analysis by the requirement $m_{\ell\ell}>20$~GeV. This confirms that the density matrix reconstructed from quantum tomography, $\rho$, using LO events, is physical to a very good approximation in the considered kinematic regime.

%------------
\subsection{NLO QCD Corrections}
\label{sec:NLOQCD}

%-----
\begin{figure}[!tb]
    \centering
   \includegraphics[width=0.32\textwidth]{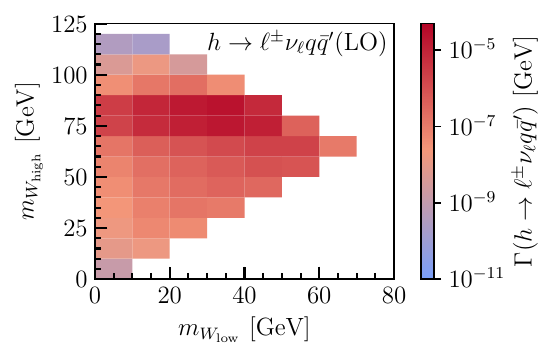}
   \includegraphics[width=0.32\textwidth]{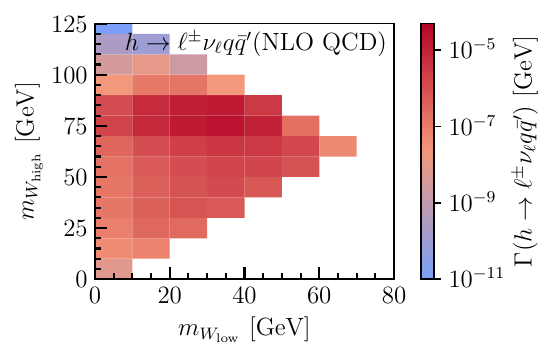}
   \includegraphics[width=0.32\textwidth]{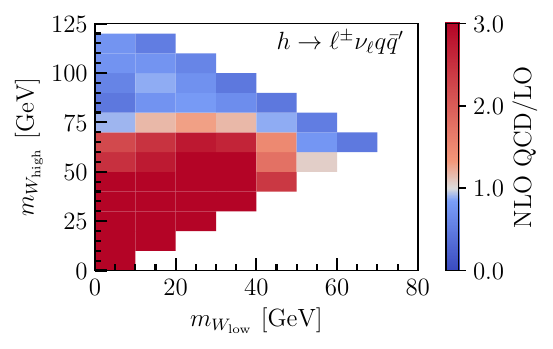} \\
   \includegraphics[width=0.32\textwidth]{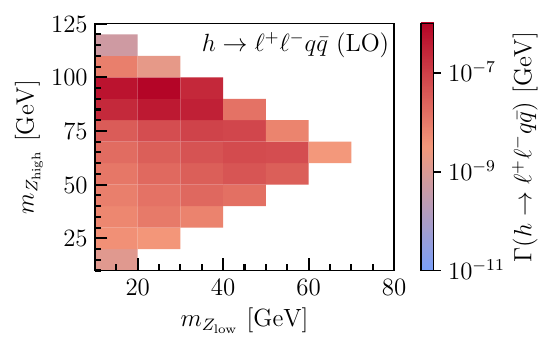}
   \includegraphics[width=0.32\textwidth]{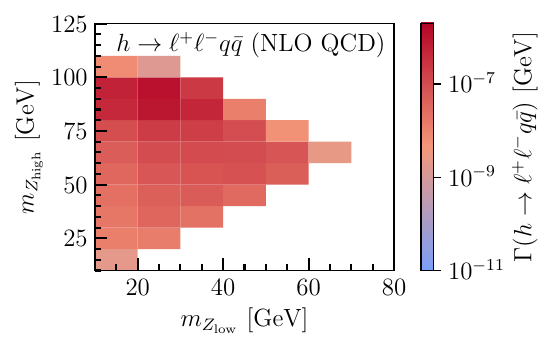}
   \includegraphics[width=0.32\textwidth]{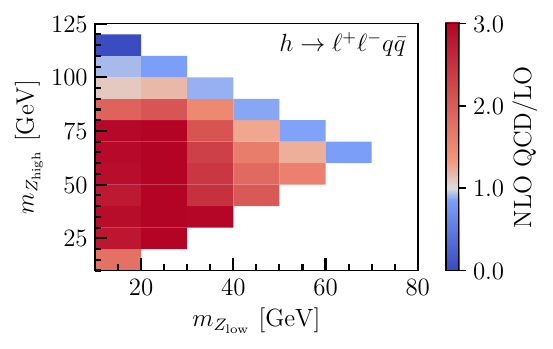} 
    \caption{Top: Differential decay distribution for $h \to \ell^\pm \nu_\ell q\bar{q}'$, with $\ell=e,\mu$ and $q,q'=\{c,s\}$, shown at LO (left), NLO QCD (center), and the corresponding NLO-QCD/LO ratio (right) in the $(m_{W_{\rm low}},m_{W_{\rm high}})$ plane. Bottom: Differential decay distribution for $h \to \ell^+\ell^- q\bar{q}$ with $q=\{c,b\}$, shown at LO (left), NLO QCD (center), and the corresponding NLO-QCD/LO ratio (right) in the $(m_{Z_{\rm low}},m_{Z_{\rm high}})$ plane. In the NLO analysis, jets are reclustered with radius parameter $R=1$.}
    \label{fig:Width-QCD}
\end{figure}
%-----

We study the NLO QCD effects in the $h\to \ell^\pm\nu_\ell q\bar q'$ and $h\to \ell^+\ell^- q\bar q$ channels, using
\texttt{MadGraph5\_aMC@NLO}~\cite{Alwall:2014hca,Frederix:2018nkq} at fixed order, and adopting the input parameters given in \Cref{eq:CMparams1,eq:CMparams2,eq:GFparam,eq:mh,eq:mt}. To quantify the impact of real radiation on event reconstruction, jets are clustered with \texttt{FastJet}~\cite{Cacciari:2011ma}, using the anti-$k_T$ algorithm with two scenarios of radius parameters: $R=0.4$ and $R=1$~\cite{Cacciari:2008gp}. Jet flavor is assigned by matching each jet to the underlying $q$ or $\bar q$ parton through an angular criterion. Jets matched to the $q$ and $\bar q$ define the directions entering the angular analysis. We require events to contain at least two jets, with the $q$ and $\bar q$ each matched to distinct jets.

In~\autoref{fig:Width-QCD}, we show the partial widths at LO (left), NLO QCD (center), and their ratios (right) in the $(m_{V_{\rm low}},m_{V_{\rm high}})$ plane, corresponding to the lower and higher invariant mass vector bosons, for $h\to \ell^\pm\nu_\ell q\bar{q}'$ (top) and $h\to \ell^+\ell^- q\bar{q}$ (bottom). The dominant contribution at both LO and NLO arises from the region in which $m_{V_{\rm high}}$ is close to the on-shell pole. Although the NLO QCD corrections can be locally sizable, as illustrated by the ratio plots, the inclusive correction to the decay width remains modest. We find 
$\delta\Gamma_\text{NLO}/\Gamma_\text{LO}\sim 0.76\%$ for $h\to \ell^\pm\nu_\ell q\bar{q}'$ and $\sim 2.78\%$ for $h\to \ell^+\ell^- q\bar{q}$, imposing $m_{\ell\ell} > 10~\text{GeV}$ and $m_{q\bar{q}} > 10~\text{GeV}$ in the latter channel. Jets are reclustered with radius $R = 1$.

%-----
\begin{table}[!tb]
\centering
\renewcommand{\arraystretch}{1.1}
\begin{tabular}{c|c|c|c|c|c}
\toprule
\multicolumn{6}{c}{$h\to \ell^\pm\nu_\ell q\bar{q}'$} \\
\midrule
& LO 
& NLO$_{R=0.4}$ 
& NLO$_{R=0.4}$/LO 
& NLO$_{R=1}$ 
& NLO$_{R=1}$/LO \\ \hline
$A^{1}_{2,0}$ & $-0.564(1)$ & $-0.5433(8)$ & $0.963(2)$ & $-0.556(4)$ & $0.986(7)$ \\
$A^{2}_{2,0}$ & $-0.564(1)$ & $-0.5411(8)$ & $0.959(2)$ & $-0.555(3)$ & $0.984(6)$ \\
$C_{1,1,1,-1}$ & $0.9560(9)$ & $0.946(4)$ & $0.989(4)$ & $0.948(2)$ & $0.991(2)$ \\
$C_{2,1,2,-1}$ & $-0.952(3)$ & $-0.95(2)$ & $1.00(2)$ & $-0.94(1)$ & $0.99(1)$ \\
$C_{1,0,1,0}$ & $-0.6035(9)$ & $-0.607(4)$ & $1.006(7)$ & $-0.602(2)$ & $0.997(4)$ \\
$C_{2,0,2,0}$ & $1.396(4)$ & $1.37(2)$ & $0.98(1)$ & $1.38(1)$ & $0.988(8)$ \\
$C_{2,2,2,-2}$ & $0.607(4)$ & $0.59(2)$ & 0.97(3) & $0.59(1)$ & $0.97(2)$ \\
$A^{1}_{2,1}$ & $0$ & $0.07(1)$ & -- & $0.054(7)$ & -- \\
$A^{1}_{2,2}$ & $0$ & $-0.08(1)$ & -- & $-0.066(8)$ & -- \\
$A^{2}_{2,1}$ & $0$ & $0.07(1)$ & -- & $0.055(7)$ & -- \\
$A^{2}_{2,2}$ & $0$ & $-0.08(1)$ & -- & $-0.066(8)$ & -- \\
$C_{1,1,1,1}$ & $0$ & $0.05(1)$ & -- & $0.037(5)$ & -- \\
$C_{1,1,1,0}$ & $0$ & $-0.026(6)$ & -- & $-0.020(3)$ & -- \\
$C_{1,0,1,1}$ & $0$ & $-0.025(6)$ & -- & $-0.019(3)$ & -- \\
\bottomrule
\end{tabular}
\caption{Angular coefficients for $h\to \ell^\pm\nu_\ell q\bar q'$ at LO, NLO QCD, and the corresponding NLO/LO ratios. The NLO results are shown for two jet radii, $R=0.4$ and $R=1$. We impose the mass window $|m_{q\bar q'} - m_W| < 10~\text{GeV}$ to select the on-shell region of the hadronically decaying $W$ boson. The combined statistical and theoretical uncertainties are shown in parentheses. Theoretical uncertainties are estimated by varying the renormalization scale between $\mu_R/2$ and $2\mu_R$.}
\label{tab:WWQCD}
\end{table}
%-----

%-----
\begin{table}[!t]
    \centering
\begin{tabular}{c|c|c|c|c|c}
\toprule
 & \multicolumn{5}{c}{$h\to \ell^+\ell^- c\bar{c}$} \\
\midrule
             & LO & NLO$_{R=0.4}$ & NLO$_{R=0.4}$/LO & NLO$_{R=1}$ & NLO$_{R=1}$/LO \\
\hline
$A_{2,0}^1$        & $-0.329(1)$ & $-0.316(8)$ & $0.96(2)$ & $-0.326(4)$ & $0.99(1)$ \\
$A_{2,0}^2$        & $-0.328(1)$ & $-0.29(1)$ & $0.88(3)$ & $-0.315(4)$ & $0.96(2)$ \\
$C_{1,1,1,-1}$     & $1.047(4)$  & $1.07(3)$  & $1.02(3)$ & $1.07(2)$   & $1.02(2)$ \\
$C_{2,1,2,-1}$     & $-1.041(3)$ & $-1.05(2)$ & $1.01(2)$ & $-1.04(1)$  & $1.00(1)$ \\
$C_{1,0,1,0}$      & $-0.766(4)$ & $-0.81(3)$ & $1.06(4)$  & $-0.79(1)$  & $1.03(1)$ \\
$C_{2,0,2,0}$      & $1.233(3)$  & $1.22(2)$  & $0.99(2)$ & $1.22(1)$   & $0.989(8)$ \\
$C_{2,2,2,-2}$     & $0.764(3)$  & $0.76(2)$  & $0.99(2)$ & $0.76(1)$   & $0.99(1)$ \\
$A_{2,1}^{2}$      & $0$         & $0.19(3)$  & --        & $0.14(2)$  & -- \\
$A_{2,2}^{2}$      & $0$         & $-0.21(3)$ & --        & $-0.16(2)$ & -- \\
$C_{2,0,2,2}$      & $0$         & $0.11(3)$  & --        & $0.08(2)$   & -- \\
\midrule
 & \multicolumn{5}{c}{$h\to \ell^+\ell^- b\bar{b}$} \\
\midrule
             & LO & NLO$_{R=0.4}$ & NLO$_{R=0.4}$/LO & NLO$_{R=1}$ & NLO$_{R=1}$/LO \\
\hline
$A_{2,0}^1$        & $-0.327(1)$ & $-0.314(8)$ & $0.96(2)$ & $-0.321(5)$ & $0.98(1)$ \\
$A_{2,0}^2$        & $-0.327(1)$ & $-0.294(7)$ & $0.90(2)$ & $-0.313(4)$ & $0.96(1)$ \\
$C_{1,1,1,-1}$     & $1.046(3)$  & $1.04(2)$  & $0.99(2)$ & $1.03(2)$   & $0.98(2)$ \\
$C_{2,1,2,-1}$     & $-1.046(3)$ & $-1.05(3)$ & $1.00(3)$ & $-1.04(2)$  & $0.99(2)$ \\
$C_{1,0,1,0}$      & $-0.774(3)$ & $-0.80(2)$ & $1.03(3)$ & $-0.78(1)$  & $1.01(1)$ \\
$C_{2,0,2,0}$      & $1.228(4)$  & $1.22(2)$  & $0.99(2)$ & $1.22(1)$   & $0.993(9)$ \\
$C_{2,2,2,-2}$     & $0.770(3)$  & $0.77(2)$  & $1.00(2)$  & $0.76(2)$   & $0.99(3)$ \\
$A_{2,1}^{2}$      & $0$         & $0.06(1)$  & --        & $0.09(1)$  & -- \\
$A_{2,2}^{2}$      & $0$         & $-0.07(1)$ & --        & $-0.10(1)$ & -- \\
$C_{2,0,2,2}$      & $0$         & $0.03(2)$  & --        & $0.05(2)$   & -- \\
\bottomrule
\end{tabular}
    \caption{Angular coefficients for $h \to \ell^+\ell^- q\bar q$ at LO and NLO QCD, together with the corresponding NLO/LO ratios. We impose the following kinematic selections: $|m_{q\bar q}-m_Z|<10~\text{GeV}$ and $m_{\ell\ell}>20\,\rm{GeV}$. The combined statistical and theoretical uncertainties are shown in parentheses.}
    \label{tab:ZZQCD}
\end{table}
%-----

We then perform the QT analysis described in~\autoref{sec:tomography} to assess whether these small corrections persist at the level of angular observables. The nonvanishing angular coefficients are presented in \Cref{tab:WWQCD,tab:ZZQCD}, imposing the invariant mass selection $|m_{q\bar{q}^{(\prime)}}-m_V|<10~\text{GeV}$ for the hadronic system. For the $h\to \ell^+\ell^-q\bar{q}$ channel, we additionally require $m_{\ell\ell}>20$~GeV. These selections follow the findings of \autoref{subsec:LOdecay}, suppressing contributions that arise from quark mass effects, which cannot be described within the two-qutrit state. For the $\ell^\pm\nu_\ell q\bar q'$ channel in \autoref{tab:WWQCD}, the QCD  corrections to the nonzero LO coefficients remain below $\mathcal{O}(4\%)$ for $R=0.4$. Increasing the jet radius to $R=1$ systematically reduces these effects to $\mathcal{O}(2\%)$. A larger jet radius captures more final state radiation, improving the stability of the reconstructed dijet system and the inferred quark axes that enter the angular basis. A similar pattern is observed for the $h\to \ell^+\ell^- q\bar{q}$ channel in \autoref{tab:ZZQCD}, where corrections reach up to $\mathcal{O}(10\%)$ for $R=0.4$ and are suppressed to $\mathcal{O}(5\%)$ for $R=1$. The largest effects appear in the polarization coefficients $A^i_{LM}$ with $i=2$ in the $h\to ZZ^*$ channel, where this index corresponds to the hadronically decaying boson. In contrast, in the $h\to WW^*$ channel either boson can decay hadronically, with $i=1$ is associated with the $W^+$ boson. Hence, higher-order corrections are shared between $i=1$ and $i=2$ in the $WW^*$ channel, leading to milder effects. For further details on these conventions, see \autoref{sec:tomography}.

In \autoref{tab:ZZQCD}, we also observe that the dominant angular coefficients for
$h\to \ell^+\ell^- c\bar c$ and $h\to \ell^+\ell^- b\bar b$ are numerically very similar at both LO and NLO. This behavior is expected, since the kinematic selection $|m_{q\bar q}-m_Z|<10~\text{GeV}$ keeps the hadronic system close to the on-shell
$Z$ pole, where $x_q^2=(m_q/m_{q\bar q})^2\ll 1$ and quark mass effects are strongly
suppressed. This similarity persists at NLO for the coefficients already present at LO,
consistent with the vectorlike and flavor universal structure of QCD in the
massless quark limit.
Coefficients that vanish at LO can, however, be generated at NLO due to the breakdown of exact back-to-back kinematics in the two-jet system induced by additional radiation. These effects lead to small but nonzero values for otherwise vanishing moments.

\begin{figure}[!tb]
    \centering
   \includegraphics[width=0.325\textwidth]{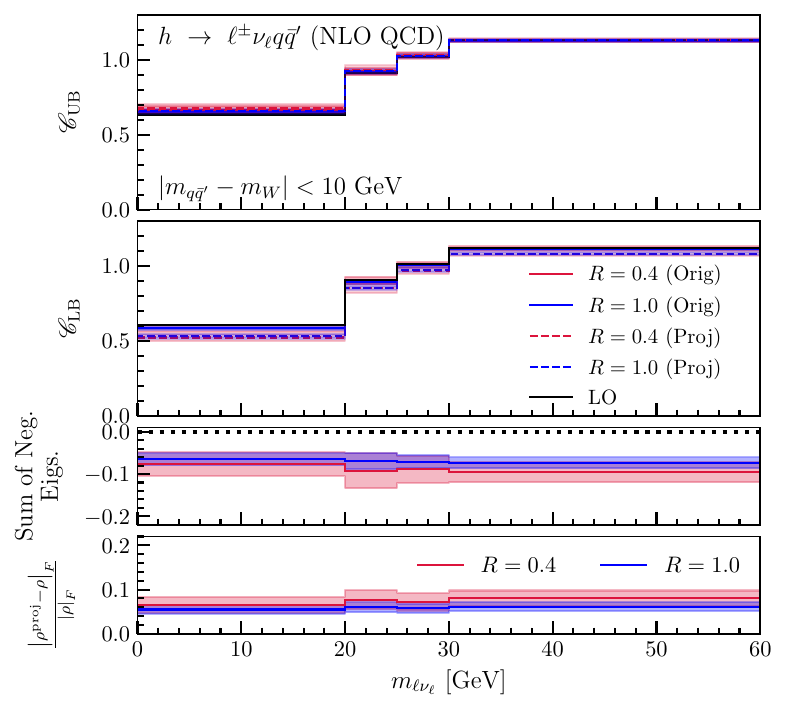}   
   \includegraphics[width=0.325\textwidth]{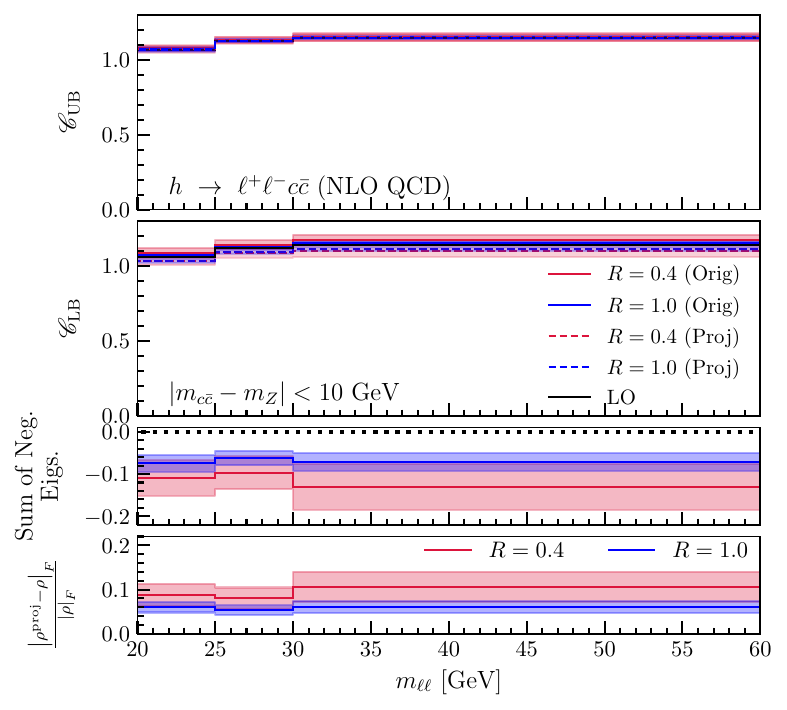}
   \includegraphics[width=0.325\textwidth]{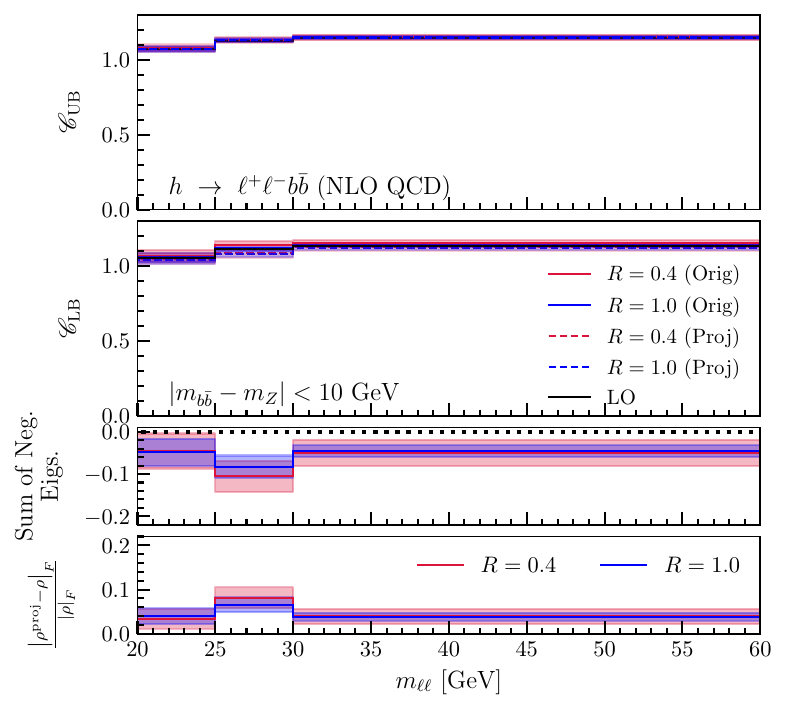} 
    \caption{Upper ($\mathcal{C}_{\rm UB}$, top subpanel) and lower ($\mathcal{C}_{\rm LB}$, second subpanel) bounds on the concurrence as functions of the dilepton invariant mass $m_{\ell\nu_\ell}$ or $m_{\ell\ell}$. The NLO QCD analysis considers $h\to \ell^\pm \nu_\ell q\bar{q}'$ (left panel), $h\to \ell^+\ell^- c\bar{c}$ (center panel), and $h\to \ell^+\ell^- b\bar{b}$ (right panel). Results are shown at LO (black) and NLO QCD for two jet radii, $R=0.4$ (red) and $R=1$ (blue). At NLO, we compare results obtained with the reconstructed density matrix $\rho$ (solid lines) to those from its closest physical projection, $\rho^\text{proj}$ (dashed lines), defined using the Frobenius norm. The shaded bands represent statistical and theoretical uncertainties, with the latter estimated by varying the renormalization scale between $\mu_R/2$ and $2\mu_R$. The third subpanel in each panel displays the sum of negative eigenvalues, while the bottom subpanel shows the relative distance $|\rho^\text{proj}-\rho|_F/|\rho|_F$, see~\autoref{eq:Frobenius} for the definition of the Frobenius norm.} 
    \label{fig:proj_cc_bb_NLOQCD}
\end{figure}

We now turn to the entanglement indicators. In \autoref{fig:proj_cc_bb_NLOQCD}, we present the lower and upper bounds on the concurrence as functions of the invariant mass of the leptonically decaying vector boson. We find that these bounds remain stable when comparing LO and NLO results. A similar stable behavior is observed when varying the jet radius from $R=0.4$ to $R=1$, as well as when comparing the reconstructed density matrix, $\rho$, with its closest physical projection, $\rho^\text{proj}$. To assess the consistency of the reconstructed density matrix, we consider both the sum of negative eigenvalues and the relative distance, defined using the Frobenius norm $|\rho-\rho^\text{proj}|_F/|\rho|_F$. In both cases, the deviations decrease as the jet radius increases. For $R=1$, the relative distance remains below $\mathcal{O}(6\%)$ across the full range of dilepton invariant mass in all three channels, confirming that NLO QCD effects on the quantum tomography reconstruction of $\rho$ are small.

%-----
%-----

%------------
\subsection{NLO EW Corrections}
\label{sec:NLOEW}

We now assess the impact of the NLO electroweak (EW) corrections in the semi-leptonic channels $h\to \ell^\pm\nu_\ell q\bar{q}'$ and $h\to \ell^+\ell^- q\bar{q}$. In the fully leptonic final states from $h \to VV^*$ decays, these effects are known to play a qualitatively important role, distorting the angular structure and, in some cases, challenging the two-qutrit interpretation of the system~\cite{Goncalves:2025mvl,Goncalves:2025xer}. Thus, it is crucial to quantify their impact in the present semi-leptonic channels. NLO EW events are generated at fixed order with \texttt{MadGraph5\_aMC@NLO}~\cite{Alwall:2014hca,Frederix:2018nkq}, using the input parameters presented in \Cref{eq:CMparams1,eq:CMparams2,eq:GFparam,eq:mh,eq:mt}. Representative Feynman diagrams are shown in~\Cref{fig:Feyn_h_lvjj,fig:Feyn_h_2j_2j}. Photons are recombined with the closest fermion (charged leptons or quarks) within a cone radius $R=0.1$. 

\begin{figure}[!tbp]
    \centering
   \includegraphics[width=0.4\textwidth]{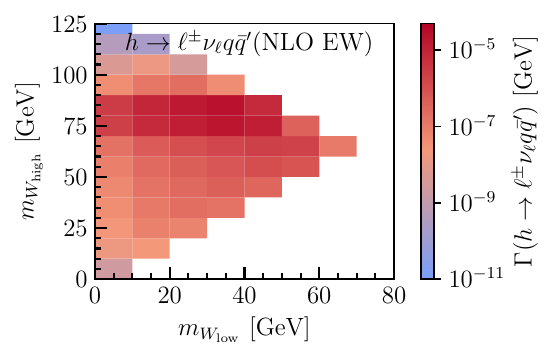}
   \includegraphics[width=0.4\textwidth]{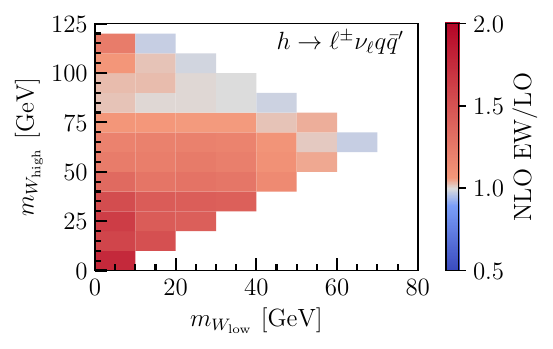} \\
   \includegraphics[width=0.4\textwidth]{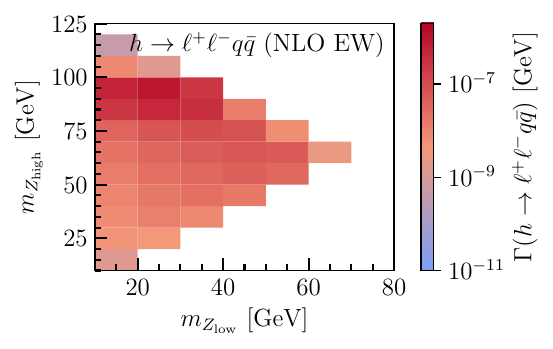}
   \includegraphics[width=0.4\textwidth]{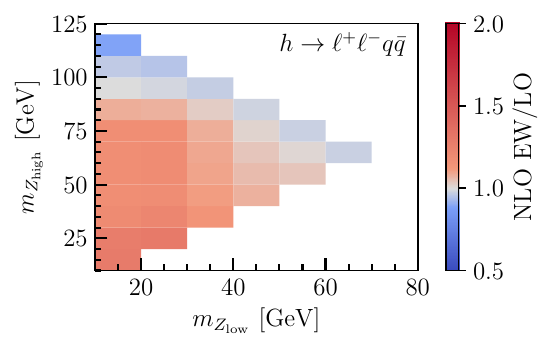}
    \caption{Top: Decay distribution for $h \rightarrow \ell^\pm \nu_\ell q\bar{q}'$ at NLO EW (left panel) and their corresponding NLO-EW/LO ratios (right panel) in the $m_{W_{\rm low}}$-$m_{W_{\rm high}}$ plane. Bottom: Decay distribution for $h \rightarrow \ell^+ \ell^- q\bar{q}$ at NLO EW (left panel), and their corresponding NLO-EW/LO ratios (right panel) in the $m_{Z_{\rm low}}$-$m_{Z_{\rm high}}$ plane.}
    \label{fig:Width-EW}
\end{figure}

In~\autoref{fig:Width-EW}, we present the partial decay widths at NLO EW and the corresponding NLO/LO ratios in the $m_{V_\text{low}}-m_{V_\text{high}}$ plane. While NLO EW effects can be sizable in some regions, as observed in the right panels, the overall effect on the partial decay widths remains moderate, with  $\delta\Gamma_\text{NLO}/\Gamma_\text{LO}\sim 3\%$ for both $h\to\ell^\pm\nu_\ell q\bar{q}'$ and  $h\to \ell^+\ell^- q\bar{q}$, imposing $m_{\ell\ell}>10$~GeV and $m_{q\bar q}>10$~GeV in the latter channel. This pattern can be understood from the fact that the dominant contributions to the partial decay widths arise from the kinematic region $m_{V_\text{high}}\sim m_V$, as shown in the left panels of~\autoref{fig:Width-EW}, where the NLO EW corrections are comparatively mild.

%---------------------------------
%---------------------------------

We now turn to the impact of NLO EW effects on the angular coefficients and the associated quantum observables. In contrast to QCD corrections, which affect only the quark lines, electroweak corrections introduce new dynamical structures. In particular, box and pentagon virtual diagrams (see diagrams 3--5 in set $(e)$ of \Cref{fig:Feyn_h_2j_2j,fig:Feyn_h_lvjj}), which do not follow the genuine $h\to VV^*$ topology, can potentially induce sizable corrections to the angular coefficients and challenge the two-qutrit description~\cite{Goncalves:2025mvl,Goncalves:2025xer,DelGratta:2025qyp,Aguilar-Saavedra:2025byk}.\footnote{As a closure test, we consider in \autoref{app:mh300} a heavier Higgs boson with $m_h=300$~GeV, for which both vector bosons can be selected near their mass shells. In this scenario, the NLO EW corrections to the dominant angular coefficients are reduced to the percent level or below.}

We first study the NLO EW corrections on the angular coefficients in the $h\to \ell^\pm\nu_\ell q\bar{q}'$ channel. Following the QT procedure described in \autoref{sec:tomography}, we present in \autoref{tab:WWEW} the coefficients that are nonvanishing at LO, together with those induced at NLO EW, imposing $|m_{q\bar{q}'}-m_W|<10~\text{GeV}$. We find that NLO EW corrections to the LO coefficients remain below $\mathcal{O}(4\%)$, consistent with the behavior observed in the $h\to e^+\nu_e \mu^-\bar\nu_\mu$ channel~\cite{Goncalves:2025xer}. Restricting the hadronic invariant mass to the near on-shell region, $|m_{q\bar{q}'}-m_W|<10~\text{GeV}$, suppresses non-resonant topologies (such as diagrams 4--5 in set $(e)$ of \autoref{fig:Feyn_h_lvjj}), leading to milder effects on the angular coefficients. The remaining corrections arise, in particular, from singly resonant loop contributions (such as diagram 3 in set $(e)$ of \autoref{fig:Feyn_h_lvjj}) and from photon radiation (\autoref{fig:Feyn_h_lvjj}~(d)). The latter is especially relevant, as it can distort the helicity basis, but its impact can be further mitigated by increasing the photon recombination radius~\cite{Goncalves:2025xer}, as confirmed in \autoref{tab:WWEW} when varying $\Delta R$ from $0.1$ to $0.3$.

%---------------------------------
\begin{table}[!tbp]
    \centering
    \begin{tabular}{c|c|c|c|c|c}
    \toprule
       &\multicolumn{5}{c}{$h\to \ell^\pm\nu_\ell q\bar{q}'$} \\
        \midrule 
        Coeff. & LO & NLO$_{\Delta R<0.1}$ & NLO$_{\Delta R<0.1}$/LO & NLO$_{\Delta R<0.3}$ & NLO$_{\Delta R<0.3}$/LO \\ \hline
        $A_{2,0}^{1}$ & $-0.564(1)$ & $-0.545(1)$ & $0.966(2)$ & $-0.548(1)$ & $0.972(2)$ \\
        $A_{2,0}^{2}$ & $-0.564(1)$ & $-0.543(2)$ & $0.963(4)$ & $-0.547(1)$ & $0.970(2)$ \\
        $C_{1,1,1,-1}$ & $0.9560(9)$ & $0.952(2)$ & $0.996(2)$ & $0.953(2)$ & $0.997(2)$ \\
        $C_{2,1,2,-1}$ & $-0.952(3)$ & $-0.956(4)$ & $1.004(5)$ & $-0.955(3)$ & $1.003(5)$ \\
        $C_{1,0,1,0}$ & $-0.6035(9)$ & $-0.611(1)$ & $1.012(2)$ & $-0.610(1)$ & $1.011(2)$ \\
        $C_{2,0,2,0}$ & $1.396(4)$ & $1.382(4)$ & $0.989(4)$ & $1.384(4)$ & $0.991(4)$ \\
        $C_{2,2,2,-2}$ & $0.607(4)$ & $0.601(5)$ & $0.99(1)$ & $0.602(4)$ & $0.99(1)$ \\
        $A_{1,0}^{1}$ & $0$ & $-0.0071(8)$ & $-$ & $-0.0052(7)$ & -- \\
        $A_{1,0}^{2}$ & $0$ & $-0.0068(8)$ & $-$ & $-0.0050(6)$ & -- \\
        $C_{2,0,1,0}$ & $0$ & $0.013(2)$ & $-$ & $0.010(2)$ & -- \\
        $C_{1,0,2,0}$ & $0$ & $0.013(2)$ & $-$ & $0.010(2)$ & -- \\
        $C_{1,-1,2,1}$ & $0$ & $-0.008(2)$ & $-$ & $-0.005(1)$ & -- \\
        $C_{2,1,1,-1}$ & $0$ & $-0.010(2)$ & $-$ & $-0.005(1)$ & -- \\ \bottomrule
    \end{tabular}
    \caption{Angular coefficients for the $h \rightarrow \ell^\pm\nu_\ell q\bar{q}'$ channel at LO and NLO EW with their corresponding NLO/LO ratios. We require $|m_{q\bar{q}'}-m_W|<10$ GeV. Photon recombination is performed with $\Delta R<0.1$ and $\Delta R<0.3$. Theoretical uncertainties from missing two-loop EW contributions are estimated by taking the square of the one-loop corrections~\cite{Freitas:2019bre}. The uncertainties quoted in parentheses include both statistical and theoretical errors propagated from the angular distributions.}   
    \label{tab:WWEW}
\end{table}
%---------------------------------

Next, we consider the $h\to \ell^+\ell^- q\bar{q}$ channel. In this case, the QT depends on the decay parameters $\eta_f$ and $\delta_f$, which in $Z$ decays are controlled by the weak mixing angle due to the interplay of left and right-handed couplings, see~\autoref{tab:spins_Z}. Because $\eta_f$ arises from the partial cancellation between these contributions, it is highly sensitive to EW corrections. Hence, using $\sin^2\theta_W^\text{LO}$ leads to an inconsistent mapping onto the $(A,C)$ coefficients once EW corrections are included~\cite{Goncalves:2025mvl,DelGratta:2025xjp}.  To account for this, we replace $\sin^2\theta_W^\text{LO}$ in the definition of $\eta_f$ with the effective weak mixing angle $\sin^2\theta_\text{eff}^f$, as discussed in Ref.~\cite{Goncalves:2025mvl}, which absorbs the dominant radiative corrections~\cite{Sirlin:2012mh,Awramik:2006uz,Chiesa:2019nqb}.\footnote{Replacing $\sin\theta_W^{\rm LO}$ by the flavor-dependent effective mixing angle,
$\sin\theta^f_{\rm eff}$, in the definition of $\eta_f$ absorbs the universal
(oblique) corrections together with the flavor-specific $Zf\bar f$ vertex corrections
into an effective coupling, capturing only a subset of the NLO electroweak effects.
Contributions from diagrams, such as the box and pentagon topologies shown in
\autoref{fig:Feyn_h_lvjj}~(e) and \autoref{fig:Feyn_h_2j_2j}~(e), are not of the
form of a single-vertex correction and therefore cannot be reabsorbed into $\eta_f$.}

%---------------------------------
\begin{table}[!tb]
    \centering
    \begin{tabular}{c|c|c|c!{\vrule width .85pt}c|c|c}
    \toprule
         & $\sin^2\theta_{W}^\text{LO}$ & $\sin^2\theta_\text{eff}^f$ & $\sin^2\theta_\text{eff}^f/\sin^2\theta_{W}^\text{LO}$ & $\eta_f^\text{LO}$ & $\eta_f^\text{NLO}$ & $\eta_f^\text{NLO}/\eta_f^\text{LO}$  \\ \hline
       $\ell^-$ & \multirow{3}{*}{$0.22289$} & $0.23230$ & $1.0422$ & $0.21430$ & $0.14088$ & $0.6574$ \\ 
       $c$ & & $0.23220$ & $1.0418$ & $0.69661$ & $0.66514$ & $0.9548$  \\ 
       $b$ & & $0.23367$ & $1.0484$ & $0.94087$ & $0.93414$ & $0.9928$  \\ \bottomrule
    \end{tabular}
    \caption{Values of the weak mixing angle at LO ($\sin^2\theta_W^\text{LO}$), at one-loop ($\sin^2\theta_\text{eff}^f$), and their ratios for charged leptons $\ell=e,\mu$, and $c$ and $b$-type quarks. For completeness, we also provide the numerical values of the spin analyzing powers $\eta_f$. We consider the same input parameters as in \Cref{eq:CMparams1,eq:CMparams2,eq:GFparam,eq:mh,eq:mt}. The effective angles are computed at one-loop with GRIFFIN package~\cite{Chen:2022dow} in the $(G_\mu,m_W,m_Z)$ input scheme.} 
    \label{tab:etaNLO}
\end{table}
%---------------------------------

We then perform the QT of the $\ell^+\ell^- q\bar{q}$ system. The results for the angular coefficients are presented in~\autoref{tab:ZZEW}. We find that the NLO EW effects can be sizable for the coefficients with $L=1$, such as $C_{1,0,1,0}$ and $C_{1,1,1,-1}$, whereas they are more modest for the other coefficients. While these general features align with previous results for the $h\to e^+e^-\mu^+\mu^-$ channel~\cite{Goncalves:2025mvl,DelGratta:2025qyp,Grossi:2024jae,DelGratta:2025xjp}, the overall impact of the NLO EW on the coefficients is milder in the semi-leptonic decay $h\to \ell^+ \ell^- q \bar{q}$. In particular, the sign flip observed in the coefficients $C_{1,0,1,0}$ and  $C_{1,1,1,-1}$ for fully leptonic states~\cite{Goncalves:2025mvl} is not present in this case. This can be attributed to the reduced effect of EW corrections in the presence of hadronic final states, which, in particular, leads to a more stable spin analyzing power (see~\autoref{tab:etaNLO}) and suppressed sensitivity to additional EW contributions beyond the $ZZ^*$ topology.

%---------------------------------
\begin{table}[!tbh]
    \centering
    \begin{tabular}{c|c|c|c|c|c}
\toprule
 & \multicolumn{5}{c}{$h\to \ell^+\ell^- c\bar{c}$} \\
\midrule
             & LO & NLO$_{\Delta R<0.1}$ & NLO$_{\Delta R<0.1}$/LO & NLO$_{\Delta R<0.3}$ & NLO$_{\Delta R<0.3}$/LO \\
\hline
$A_{2,0}^1$        & $-0.329(1)$ & $-0.319(5)$ & $0.97(1)$ & $-0.324(5)$ & $0.98(1)$ \\
$A_{2,0}^2$        & $-0.328(1)$ & $-0.327(6)$ & $0.99(2)$ & $-0.329(6)$ & $1.00(2)$ \\
$C_{1,1,1,-1}$     & $1.047(4)$ & $0.48(1)$ & $0.46(1)$ & $0.50(1)$ & $0.48(1)$ \\
$C_{2,1,2,-1}$      & $-1.041(3)$ & $-1.034(8)$ & $0.993(8)$ & $-1.033(7)$ & $0.992(7)$ \\
$C_{1,0,1,0}$     & $-0.766(4)$ & $-0.25(3)$ & $0.32(4)$ & $-0.26(2)$ & $0.34(3)$  \\
$C_{2,0,2,0}$      & $1.233(3)$ & $1.22(2)$ & $0.99(2)$ & $1.22(1)$ & $0.989(8)$ \\
$C_{2,2,2,-2}$     & $0.764(3)$ & $0.768(7)$ & $1.005(9)$ & $0.766(5)$ & $1.003(7)$ \\
\midrule
 & \multicolumn{5}{c}{$h\to \ell^+\ell^- b\bar{b}$} \\
\midrule
             & LO & NLO$_{\Delta R<0.1}$ & NLO$_{\Delta R<0.1}$/LO & NLO$_{\Delta R<0.3}$ & NLO$_{\Delta R<0.3}$/LO \\
\hline
$A_{2,0}^1$        & $-0.327(1)$ & $-0.31(1)$ & $0.95(3)$ & $-0.32(1)$ & $0.98(3)$ \\
$A_{2,0}^2$        & $-0.327(1)$ & $-0.32(1)$ & $0.98(3)$ & $-0.33(1)$ & $1.01(3)$ \\
$C_{1,1,1,-1}$     & $1.046(3)$ & $0.52(1)$ & $0.50(1)$ & $0.527(9)$ & $0.504(9)$ \\
$C_{2,1,2,-1}$      & $-1.046(3)$ & $-1.04(1)$ & $0.99(1)$ & $-1.04(1)$ & $0.99(1)$ \\
$C_{1,0,1,0}$     & $-0.774(3)$ & $-0.27(3)$ & $0.35(4)$ & $-0.274(9)$ & $0.35(1)$ \\
$C_{2,0,2,0}$      & $1.228(4)$ & $1.23(3)$ & $1.00(1)$ & $1.23(3)$ & $1.00(2)$ \\
$C_{2,2,2,-2}$     & $0.770(3)$ & $0.753(6)$ & $0.978(9)$ & $0.753(5)$ & $0.978(8)$ \\
\bottomrule
\end{tabular}
    \caption{Angular coefficients for the $h \to \ell^+\ell^- q\bar q$ channel at LO and NLO EW, shown separately for $q=c$ and $q=b$, together with the corresponding NLO/LO ratios. We require $|m_{q\bar q}-m_Z|<10$~GeV and $m_{\ell\ell}>20$~GeV. Photon recombination is performed using two cone sizes scenarios, $ \Delta R<0.1$ and $\Delta R<0.3$. The uncertainties quoted in parentheses combine statistical and theoretical contributions.}
    \label{tab:ZZEW}
\end{table}
%---------------------------------

%---------------------------------
\begin{figure}[!tbp]
    \centering
   \includegraphics[width=0.32\textwidth]{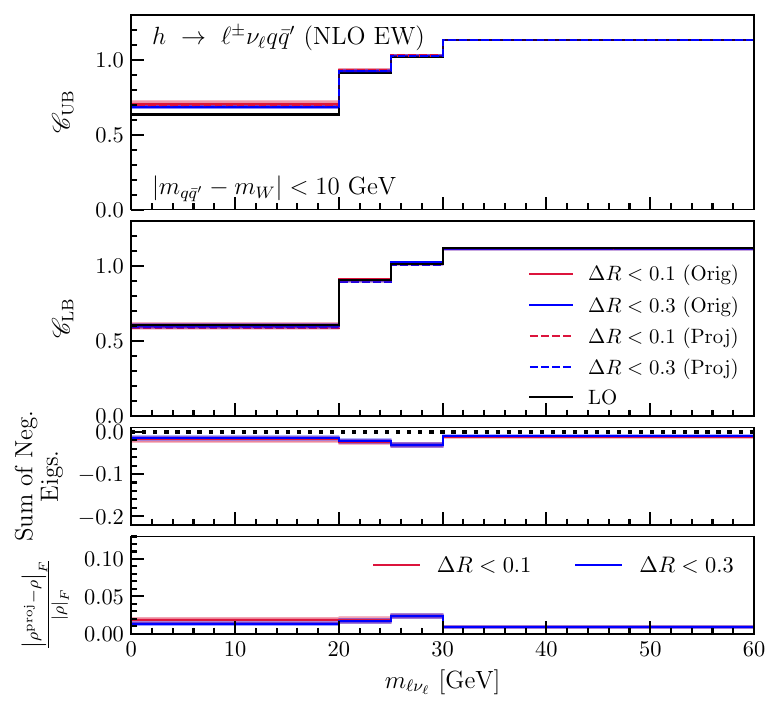}
   \includegraphics[width=0.32\textwidth]{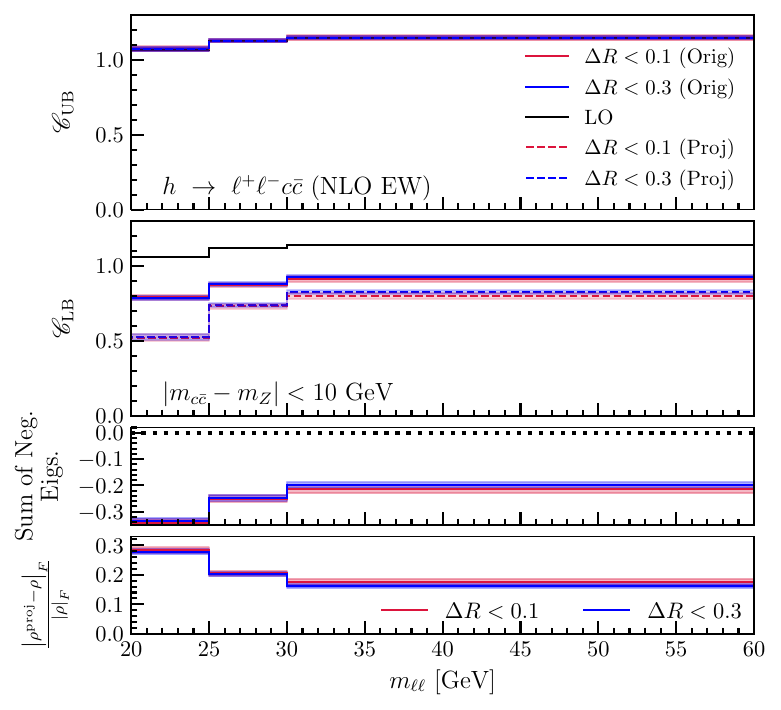}
   \includegraphics[width=0.32\textwidth]{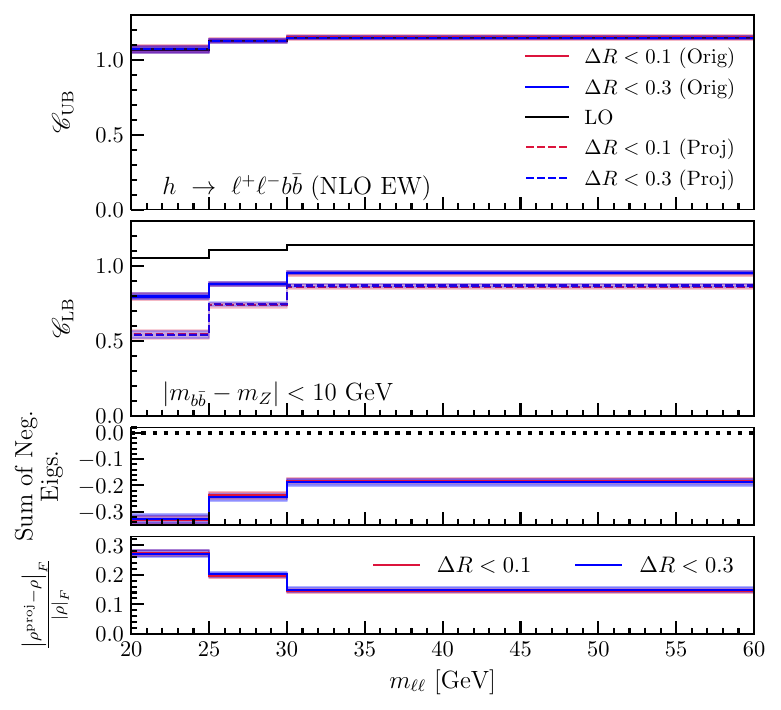} 
    \caption{Upper ($\mathcal{C}_{\rm UB}$, top subpanel) and lower ($\mathcal{C}_{\rm LB}$, second subpanel) bounds on the concurrence as functions of the dilepton invariant mass,  $m_{\ell\nu_\ell}$ or $m_{\ell\ell}$. The NLO EW analysis considers $h\to \ell^\pm \nu_\ell q\bar{q}'$ (left panel), $h\to \ell^+\ell^- c\bar{c}$ (center panel), and $h\to \ell^+\ell^- b\bar{b}$ (right panel). Results are shown at LO (black) and NLO EW for two recombination radii, $\Delta R<0.1$ (red) and $\Delta R<0.3$ (blue). The error bands account for statistical and theoretical errors due to the missing two-loop contribution (see main text for more details). The third subpanel in each figure displays the sum of negative eigenvalues, while the bottom subpanel shows the relative distance between the reconstructed spin density matrix, $\rho$, and the closest positive semi-definite, Hermitian, and unit-trace matrix, $\rho^\text{proj}$, evaluated using the Frobenius norm.}
    \label{fig:NLOEW}
\end{figure}
%---------------------------------

We now turn to the entanglement indicators and the stability of the two-qutrit description for the reconstructed spin density matrix. In \autoref{fig:NLOEW}, we present the upper $(\mathcal{C}_\text{UB})$  and lower $(\mathcal{C}_\text{LB})$ bounds on the concurrence as a function of the dilepton invariant mass. In addition to the reconstructed density matrix, $\rho$, obtained from the quantum tomography analysis described in~\autoref{sec:tomography}, we also present results using its closest physical projection, $\rho^\text{proj}$. We observe that the upper and lower bounds for the concurrence are very stable for the $h\to \ell^\pm \nu_\ell q\bar{q}'$ channel (\autoref{fig:NLOEW}, left panel) when comparing the original and projected density matrices. This is further supported by the modest negative eigenvalues for the original density matrices (third subpanel) and the small relative distances between the original and projected density matrices (bottom subpanel). For completeness, we show results for two scenarios of photon recombination radii, $R=0.1\,,0.3$. We find that the relative distances slightly decrease for larger recombination radius, as the photon recombination partially recovers the back-to-back kinematics of the LO events. This behavior is consistent with the results in~\autoref{tab:WWEW}, where the additional angular coefficients induced at NLO EW are accordingly suppressed.\footnote{Increasing the photon recombination radius leads only to subleading effects in the $h \to \ell^+\ell^- q\bar{q}$ channel, where no statistically relevant additional angular coefficients are generated. These differences between the $h\to ZZ^*$ and $h\to WW^*$ channels are consistent with the results observed in the fully leptonic final states~\cite{Goncalves:2025mvl,Goncalves:2025xer}.} This confirms the robustness of the effective two-qutrit description in the semi-leptonic $h\to \ell^\pm \nu_\ell q\bar{q}'$ channel at higher orders, further supporting the corresponding experimental analysis proposed in Ref.~\cite{Fabbri:2023ncz}.

In the $h\to \ell^+\ell^- c\bar{c}$ (central panel) and $h\to \ell^+\ell^- b\bar{b}$ (right panel), the differences between the reconstructed and projected density matrices are more sizable. However, they can be suppressed by imposing a larger dilepton invariant mass cut $m_{\ell\ell}>25$~GeV, which mitigates the negative eigenvalues from  $\sim -0.3$ to $\sim-0.15$. These larger effects in the $h \to \ell^+\ell^- q\bar{q}$ channels are consistent with the behavior observed in the angular coefficients reported in \autoref{tab:ZZEW}.

Finally, it is instructive to compare these results with those obtained for the fully leptonic channel $h\to e^+e^-\mu^+\mu^-$~\cite{Goncalves:2025mvl,Goncalves:2025xer}. In that case, higher-order EW corrections significantly challenge the stability of the density matrix, yielding relative distances $|\rho^\text{proj}-\rho|_F/|\rho|_F\sim 30\%$--$70\%$ (see Appendix~A of Ref.~\cite{Goncalves:2025xer}), signaling the breakdown of the two-qutrit approximation. In contrast, the semi-leptonic $h\to \ell^+\ell^-q\bar q$ channels, shown in the central and right panels of \autoref{fig:NLOEW}, display milder deviations, with relative distances of approximately $15\%$--$30\%$, depending on the dilepton invariant mass. Therefore, the semi-leptonic density matrices are better approximated by a two-qutrit description than their fully leptonic counterpart.

%%%%%%%%%%%%%%%%
\section{Summary}
\label{sec:summary}

Higgs decays into electroweak gauge bosons, $h\to ZZ^*, WW^*$, provide a powerful probe of the Higgs sector through their angular correlations. In this work, we have presented a systematic analysis of angular observables and quantum tomography in semi-leptonic channels, $h\to \ell^+\ell^- q\bar q$ and $h\to \ell^\pm\nu_\ell q\bar q'$, extending previous studies of fully leptonic final states. At leading order and in the massless limit for the final state fermions, angular momentum and parity constrain the decay to depend on only two independent angular coefficients, leading to entanglement across the full phase space. Finite fermion masses and radiative corrections deform this structure, making their determination crucial for precision studies, new physics searches, and quantum information interpretations.

At leading order, we identified kinematic regimes where fermion mass-induced contributions beyond the two-qutrit description are enhanced. In particular, when the hadronically decaying boson is off-shell, its scalar component can couple to the massive fermion current. These effects can be significant in inclusive hadronic configurations, especially in $h\to ZZ^*\to \ell^+\ell^-b\bar b$, but are strongly suppressed by requiring the hadronic system to lie near the vector boson mass shell. In the $h\to\ell^+\ell^-q\bar q$ channels, the additional requirement $m_{\ell\ell}>20$ GeV suppresses a separate contribution in which the $q\bar q$ pair is produced directly through the charm or bottom Yukawa coupling, while the lepton pair arises from $Z/\gamma^*$ radiation off the quark line. With these selections, the two-qutrit picture holds to good accuracy.

Including higher-order effects, we find that NLO QCD corrections induce moderate shifts in the dominant angular coefficients, at the few-percent level, reaching ${\cal O}(4\%)$ in $h\to WW^*$ and ${\cal O}(10\%)$ in $h\to ZZ^*$, and are further suppressed when using larger jet radii that better capture final state radiation. In contrast, NLO electroweak corrections introduce qualitatively new effects associated with non-factorizable contributions. These remain below ${\cal O}(4\%)$ in $h\to \ell^\pm\nu_\ell q\bar q'$, but can reach ${\cal O}(70\%)$ in $h\to \ell^+\ell^- q\bar q$. This pattern is consistent with fully leptonic analyses, where the $h\to WW^*$ channel remains comparatively stable~\cite{Goncalves:2025xer}, while $h\to ZZ^*$ displays larger corrections~\cite{Goncalves:2025mvl,DelGratta:2025qyp,Grossi:2024jae,DelGratta:2025xjp}. 
The kinematic selections introduced to suppress the quark mass contributions also improve the stability of the higher-order reconstruction. At NLO EW, the near on-shell requirement on the hadronic system suppresses non-resonant box- and pentagon-diagram contributions, while increasing the lower threshold on $m_{\ell\ell}$ further reduces the negative eigenvalues for the reconstructed density matrix $\rho$ and the Frobenius distance $|\rho^{\rm proj}-\rho|_F/|\rho|_F$ in the $ZZ^*$ channels. Analogous selections improve the stability of the fully leptonic $h\to e^+e^-\mu^+\mu^-$ reconstruction at NLO EW, but remain insufficient to recover a robust effective two-qutrit description~\cite{Goncalves:2025mvl,Goncalves:2025xer}.

Looking ahead, the increased data sets at the High-Luminosity LHC will enable precise measurements of angular correlations across multiple Higgs decay channels. In this regime, angular coefficients, such as those defined through the multipole expansion presented here, are expected to play an increasingly central role as precision observables, providing a direct interface between experiment and theory and offering sensitivity to both new physics effects and quantum information observables. Our results contribute to establishing the theoretical control for such analyses in semi-leptonic final states in the $h\to VV^*$ channels.

%%%%%%%%%%%%%%%%%%%%%%%%%%%%%%%%%%%%%%%%%%%%%%%%%%%%%%%%%%%%%%%%%%%%
\section*{Acknowledgments}
%%%%%%%%%%%%%%%%%%%%%%%%%%%%%%%%%%%%%%%%%%%%%%%%%%%%%%%%%%%%%%%%%%%%
%%%%%%%%%%%%%%%%%%%%%%%%%%%% %%%%%%%%%%%%%%%%%%%%%%%%%%%%%%%%%%%%%%%%
The work of DG and AK is supported in part by US Department of Energy Grant Number DE-SC0016013. Some computing for this project was performed at the High Performance Computing Center at Oklahoma State University, supported in part through the National Science Foundation grant OAC-1531128. AK is supported by the National Natural Science Foundation of China (grant No. E4146602), and the Fundamental Research Funds for the Central Universities (grant No. E4EQ6605X2 and E5ER6601A2).
%%%%%%%%%%%%%%%%%%%%%%%%%%%%%%%%%%%%%%%%%%%%
%%%%%%%%%%%%%%%%%%%%%%%%%%%%%%%%%%%%%%%%%%%%%%%%%%%%%%%%%%%
\appendix

\section{Mass Effects and Validity of the Two-Qutrit Description}
\label{app:spin0}

In this appendix, we assess the importance of selecting the hadronically decaying vector boson close to its mass shell in order to preserve the two-qutrit description of the $h\to VV^*$ system at LO. In the main text, this choice suppresses mass effects and ensures that both bosons can be consistently treated as spin-1 states.  To demonstrate the relevance of this setup, we invert the kinematic selection by requiring the leptonic system to be near on-shell, while allowing the hadronic system to be off-shell. This configuration enhances quark mass effects in the hadronic decay. In particular, finite fermion masses induce a spin-0 component in the off-shell vector boson, leading to deviations from the LO relations in~\autoref{eq:LOConds}.

More concretely, we impose $|m_{V_{\rm lep}}-m_V|<10~\text{GeV}$ and $20~\text{GeV}<m_{V_{\rm had}}<30~\text{GeV}$, and present the corresponding angular coefficients in \autoref{tab:spin0_A_C_coeffs}. The channels $h\to \ell^+\nu_\ell \bar{c} s$ and $h\to \ell^+\ell^- c\bar c$ remain consistent with the LO relations in \autoref{eq:LOConds}, reflecting the small charm quark mass. In contrast, the $h\to \ell^+\ell^- b\bar b$ decay shows clear deviations, for instance in the relations $A^1_{2,0}=A^2_{2,0}$ and $C_{1,1,1,-1}=-C_{2,1,2,-1}$, driven by the larger bottom quark mass, which, in particular, enhances the scalar component of the off-shell $Z^*$. These results highlight that restricting the hadronically decaying vector boson to the near on-shell region is crucial for suppressing mass effects and ensuring the validity of the two-qutrit approximation at LO.

%---------------
\begin{table}[!t]
\centering
\small
\setlength{\tabcolsep}{1pt}
\resizebox{\textwidth}{!}{
\begin{tabular}{l|ccc|ccc|ccc}
\toprule
 & \multicolumn{3}{c|}{$h\to \ell^+\nu_\ell \bar{c} s$}
 & \multicolumn{3}{c|}{$h\to \ell^+\ell^- c\bar{c}$}
 & \multicolumn{3}{c}{$h\to \ell^+\ell^- b\bar{b}$} \\
\cmidrule(lr){2-4}\cmidrule(lr){5-7}\cmidrule(lr){8-10}
 & $m_c=0$ & $m_c=1.3$~GeV & Ratio
 & $m_c=0$ & $m_c=1.3$~GeV & Ratio
 & $m_b=0$ & $m_b=4.5$~GeV & Ratio \\
\midrule
$A_{2,0}^1$ & $-0.754(3)$ & $-0.752(3)$ & $0.997(5)$
            & $-0.408(2)$ & $-0.413(2)$ & $1.012(7)$
            & $-0.408(2)$ & $-0.435(2)$ & $1.066(7)$ \\

$A_{2,0}^2$ & $-0.752(3)$ & $-0.751(3)$ & $0.998(5)$
            & $-0.408(2)$ & $-0.404(2)$ & $0.990(7)$
            & $-0.409(2)$ & $-0.380(2)$ & $0.929(6)$ \\

$C_{1,0,1,0}$ & $-0.472(2)$ & $-0.472(2)$ & $1.000(5)$
              & $-0.713(8)$ & $-0.720(8)$ & $1.01(1)$
              & $-0.710(7)$ & $-0.737(7)$ & $1.04(1)$ \\

$C_{1,1,1,-1}$  & $0.975(2)$ & $0.976(2)$ & $1.001(2)$
                & $1.050(7)$ & $1.065(7)$ & $1.014(9)$
                & $1.046(6)$ & $1.075(6)$ & $1.027(8)$ \\

$C_{2,1,2,-1}$  & $-0.982(6)$ & $-0.971(6)$ & $0.989(8)$
                & $-1.042(5)$ & $-1.043(5)$ & $1.001(6)$
                & $-1.043(5)$ & $-0.984(5)$ & $0.943(6)$ \\

$C_{2,0,2,0}$   & $1.537(8)$ & $1.537(8)$ & $1.000(7)$
                & $1.292(7)$ & $1.282(7)$ & $0.992(7)$
                & $1.283(7)$ & $1.204(7)$ & $0.938(7)$ \\

$C_{2,2,2,-2}$  & $0.465(8)$ & $0.467(8)$ & $1.00(2)$
                & $0.709(6)$ & $0.705(6)$ & $0.99(1)$
                & $0.712(6)$ & $0.672(6)$ & $0.94(1)$ \\

\bottomrule
\end{tabular}
}
\caption{Angular coefficients at LO for the channels $h\to \ell^+\nu_\ell \bar{c}s$, $h\to \ell^+\ell^- c\bar c$, and $h\to \ell^+\ell^- b\bar b$. For all channels, we require $|m_{V_{\rm lep}}-m_V|<10$~GeV and $20~\text{GeV}<m_{V_{\rm had}}<30$~GeV. The coefficients are presented in the massless approximation and including finite quark mass effects. For $x_q\neq 0$, we adopt the central value of the invariant mass bin, $m_{V_{\rm had}}=25\,\text{GeV}$, so that $x_q = m_q / 25\,\text{GeV}$.}
\label{tab:spin0_A_C_coeffs}
\end{table}
%---------------

%%%%%%%%%%%%%%%%%%%%%%%%%%%%%%%%%%%%%%%%%%%%%%%%%%%%%%%%%%%
\section{Charge Tagging Effects}
\label{app:chargeeffect}

In the main text, we assumed perfect knowledge of the electric charge of the quark initiating the jet from the hadronic decay of the vector boson, allowing for a consistent choice of the fermion with definite spin analyzing power in the QT procedure. In practice, however, this requirement depends on the decay channel. For $h\to WW^*$, flavor tagging is sufficient to identify the appropriate analyzer: in $W^-\to \bar{c}s$ or $W^+\to c\bar{s}$, charm tagging efficiently distinguishes the charm quark from the strange quark, enabling the selection of the analyzer, and making explicit charge tagging unnecessary. In contrast, for $h\to ZZ^*$ in semi-leptonic final states with $Z\to b\bar{b}$ or $Z\to c\bar{c}$, flavor tagging alone does not distinguish between quark and antiquark, and charge tagging becomes necessary to consistently select the analyzer. In our study, we choose the $b$ or $c$ quark as analyzer, while selecting the corresponding antiquark would lead to identical QT results provided the sign of the spin analyzing parameter $\eta_f$ is reversed.

In realistic experimental conditions, however, identifying the electric charge of the quark initiating the jet is challenging and one may attempt to neglect this information. While the raw angular coefficients $f$ and $g$ of~\Cref{eq:coef1,eq:coef2} are well defined, their relations to the coefficients $A$ and $C$ no longer hold. If the direction of the decay product used as an analyzer is chosen incorrectly, the reconstructed angular distribution entering the QT analysis is modified. In this appendix, we explicitly show that when charge identification of the jets is neglected, the reconstructed density matrix is no longer well defined.

To illustrate this effect, consider the decay of a vector boson, $V$, into two fermions $f_1$ and $f_2$. In the $V$ rest frame, $f_1$ and $f_2$ are back-to-back, so if $f_1$ makes an angle $\theta$ with the $V$ direction, $f_2$ makes an angle $\bar \theta=\pi-\theta$. If the analyzer is chosen randomly, it effectively averages over $\cos\theta$ and $\cos\bar{\theta}=\cos(\pi-\theta)=-\cos\theta$. Hence, all contributions that are odd in $\cos\theta$ vanish, and the angular coefficients with $L=1$ in~\autoref{eq:coef1} and~\autoref{eq:coef2} are zero, while the remaining coefficients are unchanged. At LO, this implies $C_{1,0,1,0}=C_{1,-1,1,1}=C_{1,1,1,-1}=0$, breaking part of the LO relations in~\autoref{eq:LOConds} and modifying the structure of the density matrix relative to~\autoref{eq:rhoLOForm}.

Explicitly, by imposing $C_{1,0,1,0}=C_{1,1,1,-1}=C_{1,-1,1,1}=0$, the density matrix takes the form
%--------------------------------------
\begin{align}
\resizebox{0.95\textwidth}{!}{$
    \rho_{\mathrm{LO}} = \begin{pmatrix}
        \frac{1}{6} C_{2,2,2,-2} & 0 & 0 & 0 & 0 & 0 & 0 & 0 & 0 \\
         0 & 0 & 0 & -\frac{1}{6}C_{2,1,2,-1} & 0 & 0 & 0 & 0 & 0 \\
         0 & 0 & \frac{1}{6}C_{2,2,2,-2} & 0 & \frac{1}{6}C_{2,1,2,-1} & 0 & \frac{1}{3}C_{2,2,2,-2} & 0 & 0 \\
         0 & -\frac{1}{6}C_{2,1,2,-1} & 0 & 0 & 0 & 0 & 0 & 0 & 0 \\
         0 & 0 & \frac{1}{6}C_{2,1,2,-1} & 0 & 1-\frac{2}{3}C_{2,2,2,-2} & 0 & \frac{1}{6}C_{2,1,2,-1} & 0 & 0 \\
         0 & 0 & 0 & 0 & 0 & 0 & 0 & -\frac{1}{6}C_{2,1,2,-1} & 0 \\
         0 & 0 & \frac{1}{3}C_{2,2,2,-2} & 0 & \frac{1}{6}C_{2,1,2,-1} & 0 & \frac{1}{6}C_{2,2,2,-2} & 0 & 0 \\
         0 & 0 & 0 & 0 & 0 & -\frac{1}{6}C_{2,1,2,-1} & 0 & 0 & 0 \\
         0 & 0 & 0 & 0 & 0 & 0 & 0 & 0 & \frac{1}{6}C_{2,2,2,-2} \\
    \end{pmatrix}.$}
        \label{eq:rhoLOForm2}
\end{align}
%--------------------------------------
The corresponding eigenvalues are
\begin{align}
\lambda_1 &= \frac{1}{6}C_{2,1,2,-1}\,,\,\lambda_2 = -\frac{1}{6}C_{2,1,2,-1}\,,\,\lambda_3=\frac{1}{6}C_{2,2,2,-2}\,,\,\lambda_4=-\frac{1}{6}C_{2,2,2,-2}\,,\nonumber \\ 
\lambda_5&=\frac{1}{12}\left(6+\sqrt{8C_{2,1,2,-1}^2+(6-7C_{2,2,2,-2})^2}-C_{2,2,2,-2}\right)\,,\\ 
\lambda_6&=\frac{1}{12}\left(6-\sqrt{8C_{2,1,2,-1}^2+(6-7C_{2,2,2,-2})^2}-C_{2,2,2,-2}\right)\,,\nonumber
\end{align}
with $\lambda_1$, $\lambda_2$, and $\lambda_3$ each doubly degenerate. This structure implies that the density matrix has negative eigenvalues if $C_{2,1,2,-1}$ or $C_{2,2,2,-2}$ are nonzero, which is indeed the case. The magnitude of the negative eigenvalues depend on $|C_{2,1,2,-1}|$ and $|C_{2,2,2,-2}|$, showing that an incorrect choice of analyzer leads to an unphysical density matrix. 

\begin{table}[!tb]
    \centering
    \begin{tabular}{c|c|c|c}
    \toprule
       Coefficient  & $b$ & $\bar{b}$ & $b/\bar{b}$  \\ \midrule
        $A_{2,0}^1$ & $-0.332(9)$ & $-0.332(9)$ & $-0.332(9)$ \\
        $A_{2,0}^2$ & $-0.320(9)$ & $-0.320(9)$ & $-0.320(9)$ \\
        $C_{1,0,1,0}$ & $-0.73(3)$ & $-0.73(3)$ & $0.02(3)$ \\
        $C_{1,1,1,-1}$ & $1.04(2)$ & $1.04(2)$ & $0.02(2)$ \\
        $C_{2,1,2,-1}$ & $-1.04(2)$ & $-1.04(2)$ & $-1.04(2)$ \\
        $C_{2,0,2,0}$ & $1.25(3)$ & $1.25(3)$ & $1.25(3)$ \\
        $C_{2,2,2,-2}$ & $0.73(2)$ & $0.73(2)$ & $0.73(2)$ \\ \bottomrule
    \end{tabular}
    \caption{Angular coefficients for $h\to\ell^+\ell^- b\bar{b}$ at LO. The coefficients are computed choosing the $b$-quark (second column), the $\bar{b}$ (third column), and randomly choosing $b$ or $\bar{b}$ (fourth column). We require $|m_{b\bar b}-m_Z|<10$ GeV and $m_{\ell\ell}>20$ GeV as in the main text. We consider the input parameters in \Cref{eq:CMparams1,eq:CMparams2,eq:GFparam,eq:mh,eq:mt} and $m_b=4.5$~GeV.}
    \label{tab:coeffsb}
\end{table}

To verify this numerically, we compute the angular coefficients randomly choosing the direction of the $b$ or $\bar b$ quarks for $h\to \ell^+\ell^- b\bar b$ at LO. The results are shown in the last column of~\autoref{tab:coeffsb}. For comparison, we also present the angular coefficients computed by consistently selecting $b$ or $\bar b$ in the second and third columns, respectively. As anticipated, the coefficients $C_{1,M,1,-M}$ are consistent with zero when the analyzer is randomly chosen, whereas the remaining coefficients are unaffected.
Moreover, the reconstructed matrix displays three negative eigenvalues, $-0.19(1)$, $-0.16(1)$, and $-0.12(1)$. These values are well approximated by $C_{2,1,2,-1}/6\approx -0.17$ and $-C_{2,2,2,-2}/6\approx -0.12$, confirming the analytic expectations.\footnote{We compute the eigenvalues using the full density matrix, which, due to numerical uncertainties, may not exactly match the form in~\autoref{eq:rhoLOForm2}. This can slightly modify the dependence of the eigenvalues on $C_{2,1,2,-1}$ and $C_{2,2,2,-2}$. Nevertheless, the effect is small, as confirmed by the agreement of the predictions within $2\sigma$.} These results highlight the importance of correctly identifying the hadronic $V$ decay analyzer in the QT of the $VV^*$ system. This is particularly relevant for $h\to ZZ^*$, where charge tagging is required, while for $h\to WW^*$ flavor tagging is already sufficient.

%%%%%%%%%%%%%%%
\section{NLO EW Effects for a Higgs Boson with $m_h=300$~GeV}
\label{app:mh300}

%---------------------------------
\begin{table}[!h]
    \centering
    \begin{tabular}{c|c|c|c|c|c}
    \toprule
       &\multicolumn{5}{c}{$h\to \ell^\pm\nu_\ell q\bar{q}'$} \\
        \midrule 
        Coeff. & LO & NLO$_{\Delta R<0.1}$ & NLO$_{\Delta R<0.1}$/LO & NLO$_{\Delta R<0.3}$ & NLO$_{\Delta R<0.3}$/LO \\ \hline
        $A_{2,0}^{1}$ & $-1.298(1)$ & $-1.295(2)$ & $0.998(2)$ & $-1.296(2)$ & $0.998(2)$ \\
        $A_{2,0}^{2}$ & $-1.298(1)$ & $-1.294(2)$ & $0.997(2)$ & $-1.295(2)$ & $0.998(2)$ \\
        $C_{1,1,1,-1}$ & $0.4739(6)$ & $0.478(1)$ & $1.008(2)$ & $0.478(1)$ & $1.008(2)$ \\
        $C_{2,1,2,-1}$ & $-0.474(2)$ & $-0.477(3)$ & $1.006(8)$ & $-0.477(3)$ & $1.006(8)$ \\
        $C_{1,0,1,0}$ & $-0.0796(4)$ & $-0.083(7)$ & $1.04(8)$ & $-0.083(5)$ & $1.04(6)$ \\
        $C_{2,0,2,0}$ & $1.919(3)$ & $1.909(5)$ & $0.995(3)$ & $1.909(3)$ & $0.995(2)$ \\
        $C_{2,2,2,-2}$ & $0.077(3)$ & $0.083(6)$ & $1.08(9)$ & $0.083(5)$ & $1.08(7)$ \\
        \bottomrule
    \end{tabular}
    \caption{Angular coefficients for the $h \rightarrow \ell^\pm\nu_\ell q\bar{q}'$ channel at LO and NLO EW with their corresponding NLO/LO ratios for $m_h=300$~GeV. We require $|m_{V}-m_W|<10$ GeV for both hadronic and leptonically decaying W bosons. Photon recombination is performed with $\Delta R<0.1$ and $\Delta R<0.3$. Theoretical uncertainties from missing two-loop EW contributions are estimated by taking the square of the one-loop corrections~\cite{Freitas:2019bre}. The uncertainties quoted in parentheses include both statistical and theoretical errors propagated from the angular distributions.}   
    \label{tab:WWEW300}
\end{table}
%---------------------------------

%---------------------------------
\begin{table}[!htb]
    \centering
    \begin{tabular}{c|c|c|c|c|c}
\toprule
 & \multicolumn{5}{c}{$h\to \ell^+\ell^- c\bar{c}$} \\
\midrule
             & LO & NLO$_{\Delta R<0.1}$ & NLO$_{\Delta R<0.1}$/LO & NLO$_{\Delta R<0.3}$ & NLO$_{\Delta R<0.3}$/LO \\
\hline
$A_{2,0}^1$        & $-1.212(1)$ & $-1.213(3)$ & $1.001(2)$ & $-1.214(2)$ & $1.001(2)$ \\
$A_{2,0}^2$        & $-1.213(1)$ & $-1.215(3)$ & $1.002(2)$ & $-1.215(2)$ & $1.001(2)$ \\
$C_{1,1,1,-1}$     & $0.614(4)$ & $0.61(2)$ & $0.99(3)$ & $0.61(1)$ & $0.99(1)$ \\
$C_{2,1,2,-1}$      & $-0.611(2)$ & $-0.619(5)$ & $1.013(8)$ & $-0.618(4)$ & $1.011(7)$ \\
$C_{1,0,1,0}$     & $-0.137(3)$ & $-0.135(9)$ & $0.98(6)$ & $-0.134(7)$ & $0.98(5)$ \\
$C_{2,0,2,0}$      & $1.857(3)$ & $1.858(7)$ & $1.000(4)$& $1.859(5)$ & $1.001(3)$ \\
$C_{2,2,2,-2}$     & $0.144(3)$ & $0.146(8)$ & $1.01(5)$ & $0.146(6)$ & $1.01(4)$ \\
\midrule
 & \multicolumn{5}{c}{$h\to \ell^+\ell^- b\bar{b}$} \\
\midrule
             & LO & NLO$_{\Delta R<0.1}$ & NLO$_{\Delta R<0.1}$/LO & NLO$_{\Delta R<0.3}$ & NLO$_{\Delta R<0.3}$/LO \\
\hline
$A_{2,0}^1$        & $-1.211(1)$ & $-1.211(4)$ & $1.000(3)$ & $-1.212(3)$ & $1.001(3)$ \\
$A_{2,0}^2$        & $-1.212(1)$ & $-1.213(4)$ & $1.001(3)$ & $-1.214(3)$ & $1.001(2)$ \\
$C_{1,1,1,-1}$     & $0.619(3)$ & $0.61(2)$ & $0.98(3)$ & $0.61(1)$ & $0.98(2)$ \\
$C_{2,1,2,-1}$      & $-0.614(2)$ & $-0.617(7)$ & $1.00(1)$ & $-0.616(5)$ & $1.003(8)$ \\
$C_{1,0,1,0}$     & $-0.140(2)$ & $-0.143(9)$ & $1.02(6)$ & $-0.142(8)$ & $1.01(6)$ \\
$C_{2,0,2,0}$      & $1.851(3)$ & $1.848(9)$ & $0.998(5)$ & $1.848(8)$ & $0.998(5)$ \\
$C_{2,2,2,-2}$     & $0.142(3)$ & $0.14(1)$ & $0.98(7)$ & $0.139(9)$ & $0.98(6)$ \\
\bottomrule
\end{tabular}
    \caption{Angular coefficients for the $h \to \ell^+\ell^- q\bar q$ channel at LO and NLO EW for $m_h=300$~GeV, shown separately for $q=c$ and $q=b$, together with the corresponding NLO/LO ratios. We require $|m_{V}-m_Z|<10$~GeV for both hadronic and leptonically decaying Z bosons. Photon recombination is performed using two cone sizes scenarios, $ \Delta R<0.1$ and $\Delta R<0.3$. The uncertainties quoted in parentheses combine statistical and theoretical contributions.}
    \label{tab:ZZEW300}
\end{table}
%---------------------------------

The NLO EW corrections in the semi-leptonic $h\to VV^*$ channels are generally milder than those found in the fully leptonic $h\to 4\ell$ decay. Nevertheless, the results presented in \autoref{sec:NLOEW} show that they can remain sizable, particularly in the $h\to \ell^+\ell^-q\bar q$ channels. For $m_h=125$~GeV, at least one of the two vector bosons is necessarily off-shell. Therefore, it is not possible to simultaneously restrict both reconstructed bosons to their near-resonant regions, and non-doubly-resonant contributions, including box topologies, cannot be efficiently suppressed through kinematic selections.

To isolate the role of this unavoidable off-shell kinematics, we perform a closure test for a heavier Higgs boson with $m_h=300$~GeV, for which both vector bosons can be selected close to their mass shells. We consider the input parameters in \Cref{eq:CMparams1,eq:CMparams2,eq:GFparam,eq:mt}, and impose 
\begin{equation}
    |m_{V_{\rm lep}}-m_V|<10~{\rm GeV},
    \qquad
    |m_{V_{\rm had}}-m_V|<10~{\rm GeV},
\end{equation}
with $V=W$ for $h\to \ell^\pm\nu_\ell q\bar q'$ and $V=Z$ for
$h\to\ell^+\ell^-q\bar q$.

The corresponding nonvanishing angular coefficients are presented in \autoref{tab:WWEW300} and \autoref{tab:ZZEW300}. For both semi-leptonic decay classes, the dominant coefficients remain stable under NLO EW corrections, with NLO/LO ratios differing from unity at the percent level or below. This closure test supports the interpretation developed in the main text, namely that the larger NLO EW effects found for $m_h=125$~GeV are closely associated with the unavoidable off-shell kinematics of one of the vector bosons and the resulting enhanced sensitivity to non-doubly-resonant contributions.

%%%%%%%%%%%%%%%%%%%%%%%%%%%%%%%%%%%%%%%%%%%%%%%%%%%%

\end{sloppypar}
\bibliographystyle{utphys}

\bibliography{reference}
\end{document}